\documentclass[usenatbib,useAMS,usegraphicx]{mn2e}
\usepackage{mathptmx,url}

\newcommand*{\mysub}[2]{\ensuremath{#1_{\mathrm{#2}}}}
\newcommand*{\omegam}{\mysub{\Omega}{M}}
\newcommand*{\omegal}{\ensuremath{\Omega_{\Lambda}}}
\newcommand*{\satellite}[1]{\textit{#1}}
\newcommand*{\xmm}{\satellite{XMM-Newton}}
\newcommand*{\chandra}{\satellite{Chandra}}
\newcommand*{\asca}{\satellite{ASCA}}
\newcommand*{\rosat}{\satellite{ROSAT}}
\newcommand*{\einstein}{\satellite{Einstein}}
\newcommand*{\fuse}{\satellite{FUSE}}
\newcommand*{\hst}{\satellite{HST}}

\newcommand*{\prog}[1]{\textsc{#1}}
\newcommand*{\code}[1]{\texttt{#1}}

\newcommand*{\evsel}[1]{\texttt{#1}}
\newcommand*{\mekal}{\prog{mekal}}

\newcommand*{\xspec}{\prog{xspec}}

\newcommand*{\xmmpsf}{\code{xmmpsf}}
\newcommand*{\rgsxsrc}{\code{rgsxsrc}}
\newcommand*{\projct}{\code{projct}}
\newcommand*{\phabs}{\code{phabs}}
\newcommand*{\xmekal}{\code{mekal}}
\newcommand*{\vmekal}{\code{vmekal}}
\newcommand*{\mkcflow}{\code{mkcflow}}
\newcommand*{\vmcflow}{\code{vmcflow}}
\newcommand*{\ionic}[1]{\textsc{\romannumeral #1}}
\newcommand*{\ion}[2]{#1\,\ionic{#2}}
\newcommand*{\specseries}[2]{\ensuremath{\mathrm{#1}#2}}
\newcommand*{\diff}{\,\mathrm{d}}
\newcommand*{\unit}[1]{\ensuremath{\mathrm{\, #1}}}
\newcommand*{\Msun}{\unit{M_{\odot}}}
\newcommand*{\Zsun}{\unit{Z_{\odot}}}
\newcommand{\degrees}{\ensuremath{^{\circ}}}

\let\AAorig\AA
\renewcommand*{\AA}{\mbox\AAorig}

\makeatletter
\newcommand*{\sform}[1]{\ensuremath{\@sform#1ee }}

\def\@sform#1e#2e#3 {%
  \xdef\temp@sform{#2}%
  \xdef\tempA@sform{#1}%
  \ifx\temp@sform\empty
    #1
  \else
    \ifx\tempA@sform\empty
       10^{#2}
    \else
       #1 \times 10^{#2}
    \fi
  \fi
}
\makeatother

\begin{document}

\title[\xmm{} Observation of A2597]{An \xmm{} Observation of Abell 2597}
\author[R.G. Morris and A.C. Fabian]{%
  R. Glenn Morris\thanks{E-mail: gmorris@ast.cam.ac.uk} and A. C. Fabian\\
  Institute of Astronomy, Madingley Road, Cambridge CB3 0HA}
\maketitle

\begin{abstract}
  We report on a 120\unit{ks} \xmm{} observation of the galaxy cluster
  Abell~2597. Results from both the European Photon Imaging Camera
  (EPIC) and the Reflection Grating Spectrometer (RGS) are presented.
  From EPIC we obtain radial profiles of temperature, density and
  abundance, and use these to derive cooling time and entropy. We
  illustrate corrections to these profiles for projection and point
  spread function (PSF) effects. At the spatial resolution available
  to \xmm, the temperature declines by around a factor of two in the
  central 150\unit{kpc} or so in radius, and the abundance increases
  from about one-fifth to over one-half solar. The cooling time is
  less than 10\unit{Gyr} inside a radius of 130\unit{kpc}. EPIC fits
  to the central region are consistent with a cooling flow of around
  100 solar masses per year. Broad-band fits to the RGS spectra
  extracted from the central 2\unit{arcmin} are also consistent with a
  cooling flow of the same magnitude; with a preferred low-temperature
  cut-off of essentially zero. The data appear to suggest (albeit at
  low significance levels below formal detection limits) the presence
  of the important thermometer lines from \ion{Fe}{17} at 15,
  17\unit{\AA} rest wavelength, characteristic of gas at temperatures
  $\sim 0.3\unit{keV}$. The measured flux in each line is converted to
  a mass deposition estimate by comparison with a classical cooling
  flow model, and once again values at the level of 100 solar masses
  per year are obtained. These mass deposition rates, whilst lower
  than those of previous generations of X-ray observatories, are
  consistent with those obtained from UV data for this object. This
  raises the possibility of a classical cooling flow, at the level of
  around 100 solar masses per year, cooling from 4\unit{keV} by more
  than two orders of magnitude in temperature.
\end{abstract}

\begin{keywords}
  galaxies: clusters: individual: A2597 -- X-rays: galaxies: clusters
  -- cooling flows
\end{keywords}

\section{Introduction}\label{sec:intro}

Abell 2597 is a relatively nearby ($z \approx 0.08$), Abell richness
class 0 cluster of galaxies. Rich in features, it has been studied
extensively in many wavebands.

The central cD galaxy contains the radio source PMN J2323-1207 (PKS
B2322-12) \citep{wrig90,grif94}. The total flux density of this source
is $\sim 1.7\unit{Jy}$ at 1.5\unit{GHz} \citep{mcna01,owen92}, and
$\sim 0.3\unit{Jy}$ at 5.0\unit{GHz} \citep{ball93}, implying a fairly
steep spectral index $\alpha = -1.3$ ($\mysub{S}{\nu} \propto
\nu^{\alpha}$). The source is physically small, $\sim 7 \times 5, 5
\times 2 \unit{arcsec}$ at 1.5, 5.0\unit{GHz} respectively. The source
is only moderately resolved at these frequencies, but at the higher
frequency has an elongated appearance suggesting a double-lobed
structure. At 8.44\unit{GHz} \citep{sara95}, the source is resolved
into a nucleus and two lobes ($\sim 5\unit{arcsec}$ across),
orientated roughly NE--SW. A jet is clearly seen leading from the
nucleus on the SW. This jet undergoes a sharp ($\sim
90\unit{\degrees}$) bend about 0.5\unit{arcsec} from the nucleus.
Assuming a typical spectral index for the jets of $-0.7$, the inferred
lobe spectral index is $\approx -1.5$, suggestive of confinement and
synchrotron ageing. Very Long Baseline Array (VLBA) 1.3, 5.0\unit{GHz}
observations of the central 0.3\unit{arcsec} \citep{tayl99} show an
inverted spectrum ($\alpha = 0.6$) core, together with straight,
symmetric jets emanating from both sides.

Abell 2597 has previously been observed in X-rays by various
observatories, for example: \einstein{} \citep{craw89}; \rosat{}
\citep{sara95,sara97,pere98}; \asca{} \citep{whit00}; \chandra{}
\citep{mcna01}; \xmm{} \citep{stil02}. Traditionally, it has been
classified as a moderately strong cooling flow \citep[e.g.][]{fabi94}
cluster, with inferred X-ray mass deposition rate
(\unit{\Msun}\unit{yr^{-1}}): $270 \pm 41$ \citep[\rosat{}
PSPC;][$\mysub{r}{cool} = 152^{+67}_{-58}\unit{kpc}$]{pere98};
$259^{+176}_{-178}$ \citep[\asca;][]{whit00}.

A short ($\sim 20\unit{ks}$ of good time) \chandra{} observation
\citep{mcna01} highlighted the interaction between the radio source
and the X-ray gas, through the presence of so-called `ghost cavities'
of low surface brightness, seemingly coincident with spurs of low
frequency ($1.4\unit{GHz}$) radio emission. Such brightness
depressions are believed to be buoyantly rising bubbles
\citep[e.g.][]{birz04} associated with a prior ($\sim 10^{8}\unit{yr}$
previously, in this case) outburst of the central radio source. The
small physical size of the radio source means, however, that these
issues are not amenable to study with \xmm, given its $\sim
5\unit{arcsec}$ point spread function (PSF).

In the \chandra{} and \xmm{} era of high spatial and spectral
resolution, the cooling flow model is undergoing re-interpretation.
The cause is quite simply an absence (or at best a severe weakness) of
the spectral features expected from gas at the low-end ($\la
1\unit{keV}$, say; though there does not appear to be a fixed absolute
cut-off) of the X-ray temperature regime in the observed spectra of
cluster central regions. Reductions in temperature by factors of a few
\citep[e.g.][]{kaas04} with radius are seen. The expected emission
lines for several important low-temperature species (e.g.\ the
\ion{Fe}{17}{} 15 and 17\,\AA{} lines) do not seem to be present at
the levels predicted by simple models. This is a trend that appears to
be repeated in many clusters that have traditionally been thought to
harbour cooling flows \citep[e.g.][]{pete03}. Many explanations
\citep[e.g.][]{fabi01a,pete01} for this effect have been put forward;
some involving suppression of the gas cooling, some involving
departures from standard collisionally-ionized radiative cooling.
After a flirtation with thermal conduction \citep[but see
e.g.][]{soke03}, theoretical studies at present seem to be
concentrated on some form of distributed heating from a central AGN
\citep[e.g.][]{vecc04,reyn05,rusz04}, effected by buoyant plasma
bubbles.

Observations at ultra-violet (UV) wavelengths suggest that Abell 2597
may be a particularly interesting object for study in light of these
issues. Abell 2597, together with the $z = 0.06$ cluster Abell 1795,
was observed with the Far Ultraviolet Spectroscopic Explorer
\citep[\fuse;][]{moos00} mission by \citet{oege01}. \fuse{} is
sensitive to the \ion{O}{6} 1032, 1038\unit{\AA} resonance lines,
which are strong diagnostics of thermally radiating gas at
temperatures $\sim$ \sform{3e5}\unit{K}. In a traditional cooling flow
model, in which the ultimate fate of the hot gas is to cool to very
low temperatures, strong UV emission is expected in such lines as gas
cools out of the X-ray temperature regime; thereby potentially forming
a bridge between the hot, X-ray emitting gas and the cool,
\specseries{H}{\alpha} (see below) radiating gas. \ion{O}{6}
1032\unit{\AA} emission was detected in Abell 2597, with an inferred
luminosity \sform{3.6e40}\unit{erg}\unit{s^{-1}}. Using simple
cooling-flow theory, \citet{oege01} calculate the associated mass
flow-rate as 40\unit{\Msun}\unit{yr^{-1}} (within the effective radius
of the \fuse{} aperture at this redshift, $\sim 40\unit{kpc}$), for an
intermediate case between the limits of isobaric and isochoric
cooling. The \rosat{} HRI mass deposition rate of \citet{sara95},
$\approx 330\unit{\Msun}\unit{yr^{-1}}$ within a cooling radius
(defined by $\mysub{t}{cool} < 10\unit{Gyr}$) $\approx 130\unit{kpc}$,
is around three times larger (applying a simple linear $\dot{M}
\propto r$ scaling). The weakness of cluster cooling flows in
comparison to the predictions of the traditional model is thus again
demonstrated, but extended down to the UV regime. This
under-luminosity at UV temperatures may be contrasted with the
situation at \specseries{H}{\alpha} temperatures, where the Abell 2597
emission nebula (see below), like other such nebulae, is
over-luminous, even when compared to classical, high X-ray mass
deposition rates \citep{voit97}.

In contrast, Abell 1795 was not detected in \ion{O}{6}. Similarly,
\citet{leca04}, also using \fuse, were unable to detect \ion{O}{6}
emission from either of the low redshift ($z \approx 0.07$), low
Galactic column ($\mysub{N}{H} \sim \sform{4e20}\unit{cm^{-2}}$), rich
clusters Abell 2029 and Abell 3112. These results highlight the
apparently unusual (within the currently favoured non-cooling flow
paradigm) nature of Abell 2597, and motivate a more detailed X-ray
study of its properties.

The central ($\sim 15\unit{arcsec}$ in diameter) regions of Abell 2597
harbour an optical emission line nebula, as is often (and exclusively)
the case in cooling flow clusters. The $\specseries{H}{\alpha} +
[\ion{N}{2}]$ luminosity is $\sform{2.7e42}\unit{erg}\unit{s^{-1}}$,
making it among the most luminous in the sample of \citet{heck89}. The
\specseries{H}{\alpha} luminosity is many ($\sim 300$) times greater
than that expected from extrapolating the classical $\sim
10^{7}\unit{K}$ X-ray mass flow rates down to $\sim 10^{4}\unit{K}$, a
general feature of such nebulae. The discrepancy is made even worse if
the relative sizes of the nebulae and X-ray cooling regions are taken
into account. The deep optical spectra of \citet{voit97} reveal that
the Abell 2597 nebula has $T \sim 10^{4}\unit{K}$, $Z \sim
0.5\unit{\Zsun}$, $\mysub{n}{e} \sim 200\unit{cm^{-3}}$, and is
significantly reddened by (intrinsic, owing to the low Galactic
column) dust. The low inferred column density $\mysub{N}{\ion{H}{2}}
\approx \sform{3e19}\unit{cm^{-2}}$ relative to that of \ion{H}{1}
suggests that thin ionized \ion{H}{2} layers surround cold neutral
\ion{H}{1} cores. The nebula has a low ionization parameter, most
species being only singly ionized. Photoionization from hot stars is
the only ionization mechanism not ruled out by the observations, but
even the hottest O stars (in isolation) would not be able to heat the
nebula to the observed temperature. Some form of heat transfer from
the intracluster medium (ICM) may supply a comparable amount of energy
to the nebula.

The core of the central cD exhibits a significant blue optical excess
\citep[e.g.][]{mcna93,sara95}. \hst{} observations \citep{koek99}
resolve this excess into continuum emission around and across the
radio lobes, and several knots to the south-west. Line-dominated
(thereby not optical synchrotron or inverse Compton in origin) excess
appears in bright arcs around the edge of the radio lobes, and more
diffusely across the lobe faces. Such geometry is suggestive of an
emission shell surrounding the radio lobes. A bright optical knot is
also seen coincident with the southern radio hot-spot, consistent with
a massive clump responsible for the deflection of the southern jet.
The central region also shows strong dust obscuration, resolved into
several large (several hundred parsecs) regions.

In the infra-red, vibrational molecular hydrogen emission lines
characteristic of $\sim$ 1000--2000\unit{K} collisionally ionized gas
have been detected from the innermost $\sim 3\unit{arcsec}$ of the
central cluster galaxy \citep{falc98,dona00,edge02}. As with the
optical nebula, there is too much emission as compared to the
expectation from even an inordinately massive classical cooling flow,
and yet too little mass for this to be the final reservoir of a
standard, long-lived flow. Furthermore, the molecular gas is dusty,
and therefore unlikely to be the direct condensate of the hot ICM
(where dust is rapidly destroyed). Again as with the optical light,
there is a complex filamentary structure that seems to be spatially
associated with the radio source. The optical and infra-red properties
are both consistent with UV heating of the gas by a population of
young, hot stars \citep{dona00} (SFR $\sim$ few
$\unit{\Msun}\unit{yr^{-1}}$). Abell~2597 also has a possible CO
detection \citep{edge01}; indicative of very cold, molecular gas.

The Galactic column density in the direction of Abell 2597 ($l =
65.4$, $b = -64.8$) has the relatively low value
\sform{2.5e20}\unit{cm^{-2}} \citep{dick90,star92}. Recently,
\citet{barn03} have shown that the $5\sigma$ limit on any fluctuation
in the foreground \ion{H}{1} column in this direction, on scales $\sim
1\unit{arcmin}$, is \sform{0.43e20}\unit{cm^{-2}} (17 per cent).

Intrinsic \ion{H}{1} 21\unit{cm} absorption towards the central radio
source was detected by \citet{odea94}. A narrow ($\sim
200\unit{km}\unit{s^{-1}}$), redshifted ($\sim
300\unit{km}\unit{s^{-1}}$), spatially unresolved component coincident
with the nucleus is inferred to be a gas clump falling onto the cD. A
broad ($\sim 400\unit{km}\unit{s^{-1}}$), spatially extended ($\sim
3\unit{arcsec}$, i.e.\ the size of the radio source) component at the
systemic velocity is consistent with being associated with the (hence
photon bounded) \specseries{H}{\alpha}. The inferred column densities
and masses for the narrow and broad components are $\mysub{N}{H} \sim
\{8.2, 4.5\} \times 10^{20} \mysub{T}{s} \unit{cm^{-2}}$; $M \sim
\{3.5, 7\} \times 10^{7} \mysub{T}{s} \unit{\Msun}$; with \mysub{T}{s}
the spin temperature in units of 100\unit{K}. Intrinsic, redshifted
\ion{H}{1} was also seen in the VLBA observations of \citet{tayl99},
and interpreted as a turbulent, inwardly streaming, atomic torus
(scale height $\la 20\unit{pc}$) centred on the nucleus.

Our chosen cosmology has $H_{0} =
50$\unit{km}\unit{s^{-1}}\unit{Mpc^{-1}}, $\omegam = 1.0$, and
$\omegal = 0.0$. Adopting a redshift $z = 0.0822$ (see
Section~\ref{sec:rgs_redshift}), the luminosity distance is
$\mysub{D}{L} = 503\unit{Mpc}$, the angular diameter distance is
$\mysub{D}{A} = 429\unit{Mpc}$, and the angular length scale is
2.08\unit{kpc}\unit{arcsec^{-1}}. Data processing was carried out
using the \xmm{} Science Analysis
System\footnote{\url{http://xmm.vilspa.esa.es/sas/}} (SAS) version
5.4.1 (aka xmmsas\_20030110\_1802). Spectral analyses employed
\xspec\footnote{\url{http://heasarc.gsfc.nasa.gov/docs/xanadu/xspec/}}
version 11.3. Unless otherwise stated, quoted uncertainties are $1
\sigma$ (68 per cent) on a single parameter.

\section{EPIC data reduction}\label{sec:data}

Abell 2597 was observed by \xmm{} on 2003 June 27--28 (observation ID
0147330101), during revolution 0650. The nominal length of the
observation was 120\unit{ks}, unfortunately the exposure was heavily
affected by background flaring (see Fig.~\ref{fig:mos2rate}). The thin
optical blocking filter was applied to the European Photon Imaging
Camera (EPIC). The MOS \citep{turn01} and pn \citep{stru01} detectors
were both operated in Full Frame mode.

EPIC events files were regenerated from the Observation Data Files
(ODFs) using the SAS meta-tasks \code{epchain} and \code{emchain}. For
\code{emchain}, standard options were used, except that bad pixel
detection was switched on. \code{epchain} was run in two steps in
order to merge in the instrument house-keeping Good Time Interval
(GTI) data (which are otherwise ignored) before the final events list
was created.

\subsection{Background filtering and subtraction}\label{sec:backgrnd}

The EPIC background is comprised of a number of components: proton,
particle, photon. Soft protons in the Earth's magnetosphere scatter
through the mirror system, and are responsible for periods of intense
background flaring, where the instrument count rate can rise by orders
of magnitude. Such intervals can only be dealt with by excising them
from further analysis using a rate-based GTI filtering. Energetic
charged particles that pass through the detector excite fluorescent
emission lines in several of the constituent components. The strong
spatial variation of some components of this emission
(\citealt{frey02a}, \citealt{frey02b}, \citealt{lumb02}) makes it
preferable to extract a background from the same region of the
detector as the source. For the analysis of extended sources, the use
of blank-sky background templates \citep{lumb02} is therefore to be
desired.

For Abell 2597, cluster emission is detected up to around 8 and
9\unit{keV} in the MOS and pn respectively. The pn exhibits strong
fluorescence lines of Ni, Cu and Zn around 8\unit{keV}. The only
non-negligible fluorescent line visible in the pn data below
8\unit{keV} is the 1.5\unit{keV} Al line. Unlike the Cu (for example)
line, this line does not vary spatially over the detector
\citep{frey02b}, since it originates in the camera housing. Cutting
off the spectra at 7.3\unit{keV} removes very few counts, and means
that the only background line remaining in the pn is not spatially
varying. This frees us from the need to extract background spectra
from the same spatial region of the pn detector, i.e.\ enables us to
use a local background. This is desirable because it removes the
vagaries of scaling and varying cosmic background necessarily
associated with the use of a blank-sky background (see below); and it
allows us greater freedom in the filtering of background flares.
Additionally, comparison of the off-source pn spectra for the Abell
2597 and blank-sky fields reveals that the 8\unit{keV} fluorescence
lines are shifted to a lower energy by about 20\unit{eV} in the
latter. This almost certainly represents a change in calibration
between the processing of the blank-sky templates and our
re-processing of the ODFs for Abell 2597 (the effect of such a shift
would of course be largely restricted to an increase in $\chi^{2}$ in
those channels lying around the fluorescent line energies).

The only significant fluorescent lines in the MOS data are the Al and
Si lines at 1.5 and 1.7\unit{keV} respectively (the Cr, Mn, and Fe
lines around 6\unit{keV} are seen to be negligible). Both of these
features exhibit strong spatial variation \citep{lumb02} over the MOS
field of view (hereinafter fov). Using a local background for the MOS
is therefore not as viable a proposition as it is for the pn.

\begin{figure}
  \centering
  \includegraphics[angle=0,width=1.0\columnwidth]{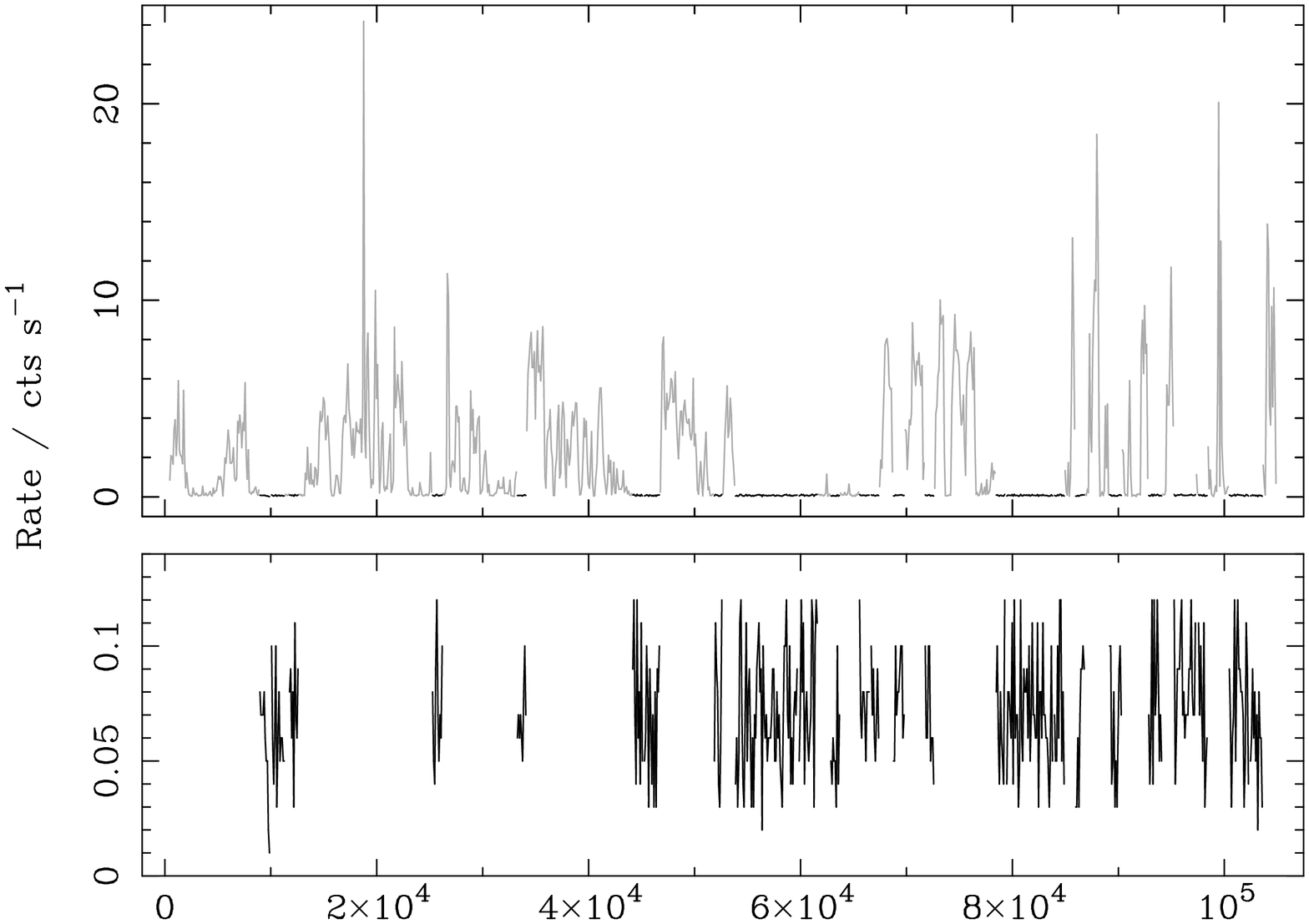}
  \caption{%
    MOS2 light curve formed from 100\unit{s} time bins. Upper panel:
    total exposure, with rejected periods of high count-rate in grey,
    accepted periods in black. Lower panel: only the selected Good
    Time Interval (GTI) is shown.%
  }
  \label{fig:mos2rate}
\end{figure}

In order to be able to make use of the blank-sky EPIC background
templates of \citet{lumb02}, we use the same form of filtering process
as employed in the construction of those files. That is, we make a
light-curve (e.g.\ Fig.~\ref{fig:mos2rate}) in 100\unit{s} time
bins of the high energy (\evsel{PI $>$ 10000}) single pixel
(\evsel{PATTERN == 0}) events across the whole detector. The standard
XMM event selection flags \evsel{\#XMMEA\_EM} and \evsel{\#XMMEA\_EP}
are applied for the MOS and pn respectively (these exclude cosmic
rays, events in bad frames, etc.).

Our preferred filtering method is to form a histogram of the number of
events in the 100\unit{s} time-bins of the light curve, fit the core
of the distribution (i.e. ignoring the high count-rate tail) with a
Gaussian, and exclude periods where the count-rate lies more than $n
\sigma$ away from the mean. Unfortunately, for MOS1 (and the pn, but
this is not an issue since we use a local background in this case) the
$2 \sigma$ count-rate thresholds lie above the filtration limits used
in the creation of the background events templates (0.2 and
0.45\unit{ct}\unit{s^{-1}} for the MOS and pn respectively). In order
to make use of the MOS1 background template we therefore apply the
same absolute count-rate filtering to this instrument as was used to
make the background events file. For MOS2, the $2 \sigma$ count-rate
threshold is 0.13\unit{ct}\unit{s^{-1}}, and this absolute value is
used to filter both the observation events file, and to re-filter the
background template events file to the same level. For the pn, the $2
\sigma$ count-rate threshold is 0.54\unit{ct}\unit{s^{-1}}.

We strengthen the filtering by requiring that all GTIs be at least 10
minutes long. The GTI files are aligned on the frame boundaries of the
relevant events files before being applied. The mean remaining ONTIME
of the events files after filtering in this way are 39 (pn), 37
(MOS1), and 38\unit{ks} (MOS2).

It is known that the particle background can vary from observation to
observation at around the ten per cent level \citep{frey02a}. We
therefore scale the products (images and spectra) of the blank-sky MOS
events file before they are used to correct the source products. The
scaling factor is obtained from the ratio of the count-rates outside
the fov, where the only events should be the result of
particle-induced fluorescence rather than from actual photons passing
through the mirror system. In detail, we find the number of counts
remaining in the GTI-filtered data after applying the \code{evselect}
filtering criterion \evsel{\#XMMEA\_16 \&\& FLAG \& 0x766a0f63 == 0
  \&\& PI in [200:12000] \&\& PATTERN $<=$ 12}, where the flag
expression selects events outside the MOS fov but with no other
non-zero flag settings. Combining this figure with the exposure time
gives an out-of-fov count-rate which is used to normalize the
background template products. In this way, we obtain scaling factors
of 1.02 and 0.91 for the MOS1 and MOS2 backgrounds respectively. To
allow for some uncertainty in the normalization, we assign a 10 per
cent systematic error to the blank-sky products, which is added in
quadrature to the statistical errors.

The SAS task \code{attcalc} was used to reproject the blank-sky
templates to the same sky position as the Abell~2597 observation, so
that the same sky coordinate selection expressions could be applied to
both the science and background fields.

Below 5\unit{keV} or so, the cosmic X-ray background (CXB) dominates
\citep{lumb02} over the internal camera background. If there were no
significant variation of the CXB over the sky, then the blank-sky
templates would also represent this background component. To allow for
any changes in the CXB in the region of the target, a so-called double
background subtraction approach can be used. The large field of view
of \xmm{}, together with the redshift of Abell 2597, result in
significant areas of the detectors being essentially free from cluster
emission. This enables us to examine the background spectrum from such
regions and compare it with those of the (scaled) blank-sky templates
extracted from the same detector region. For this purpose we extracted
spectra from an annulus lying between 7 and 10\unit{arcmin} (avoiding
regions near the edge of the fov) from the cluster centre, which we
take to be located at the emission peak, in sky coordinates $(X, Y) =
(23720.5,23720.5)$. Non-ICM sources in this region were excluded with
circular masks of radius at least 30\unit{arcsec}. The BACKSCAL FITS
header keys were corrected for the area so removed.

Initial comparison of the spectra extracted from this region of the
Abell 2597 field reveals that the agreement with the blank-sky spectra
scaled according to the out-of-fov count-rates is good; despite the
relatively poor quality of the science observation as a whole (i.e.
the heavy background flaring). There is something of a slight excess
of soft ($\la 2\unit{keV}$) counts in the Abell 2597 field, but not
to any great degree.

In summary, we use a local (extracted from 7--10\unit{arcmin} in
radius) background for the pn. For the MOS, we use the blank-sky
templates to deal with the spatially varying Si and Al background
lines, corrected by a double-background subtraction approach to
account for any (minor) variance in CXB between the science and
background fields. For the MOS, we make use of single, double, triple,
and quadruple pixel events (i.e. \evsel{PATTERN $\leqslant$ 12}). For
the pn, we use single and double events (\evsel{PATTERN $\leqslant$
  4}). In both cases only events with \evsel{FLAG == 0} are
considered.

\subsection{Out of time events}

When the pn is operated in Full Frame mode, as is the case here,
bright sources can be significantly affected by so-called out of time
(OOT) events. These are caused by photons which arrive while the CCD
is being read out (charge shifted along columns towards the readout
node). Such events are assigned a wrong RAWY coordinate and hence an
improper CTI (charge transfer inefficiency) correction. In an
uncorrected image, the OOT events due to a strong source are visible
as a bright streak smeared out in RAWY. OOT events act to broaden
spectral features.

OOT events can only be corrected for in a statistical sense. We follow
the standard procedure of re-running the SAS meta-task \code{epchain}
with the option \texttt{withoutoftime=y} passed to the \code{epevents}
task. This generates an events file that consists entirely of OOT
events, distributed over the range of RAWY present in the data. OOT
images and spectra can then be produced in the normal way, and scaled
and subtracted from source images and spectra to correct them for the
OOT events.

The appropriate scale factor is obtained from the expected fraction of
OOT events, which depends on the ratio of the readout time to the
frame time. For the pn in Full Frame mode, this is $4.6 / 73.3
\unit{ms} = 6.3$ per cent \citep[][table 1]{stru01}. A slight
correction to this factor is needed because the version of
\code{epevents} used does not take account of the fact that in pn Full
Frame mode, the first 12 (of 200) rows are marked as bad. Accordingly,
the correct scaling factor is $(1 - 12/200) * 6.3 = 5.9$ per cent. In
practice, OOT events make little difference to the pn spectral fits.

\subsection{Response files}

The lower limit for the spectral fits was placed at 500\unit{eV} for
the MOS (following the advice of \citealt{kirs03} for post-revolution
450 data; in any case the MOS results are relatively insensitive to
the adopted low-energy cut-off). For consistency (and to avoid
low-energy residuals) the same limit was applied to the pn data. The
high energy cut-off was placed at 7.3\unit{keV}, slightly below the
maximum detected energy from the cluster, to avoid the background pn
fluorescence line complex at 8\unit{keV} (as described in
Section~\ref{sec:backgrnd}).

Spectra were grouped to a minimum of 20 counts per channel to ensure
applicability of the $\chi^{2}$ statistic.

Redistribution matrix files (RMFs) and ancillary response files (ARFs)
were generated using the SAS tasks \code{rmfgen} and \code{arfgen}
respectively. RMFs were made using the standard SAS parameters. ARFs
were made in extended-source mode, using a low-resolution
detector-coordinate image of the spectral region. The image pixel size
was chosen according to the size of the extraction region so as to
adequately sample the variation in emission, bearing in mind that the
version of \code{arfgen} employed does not model variations on scales
$\la 1\unit{arcmin}$. Strictly speaking, the images should be
exposure-weighted \citep[i.e.\ de-vignetted;][]{saxt02}, but we find
this makes almost no difference in practice in this case. Bad pixel
locations were taken from the relevant events file. The approach of
generating ARFs appropriate to the emission pattern in the spectral
extraction region differs from the `weighting' method
\citep[e.g.][]{arna01} more frequently employed in \xmm{} spectral
analyses. We find that the differences in the temperature profiles
etc.\ obtained using the two approaches are negligible \citep[as
per][]{gast03}.

\section{Global properties}\label{sec:global_props}

\begin{figure}
  \centering
  \includegraphics[angle=0,width=1.0\columnwidth]{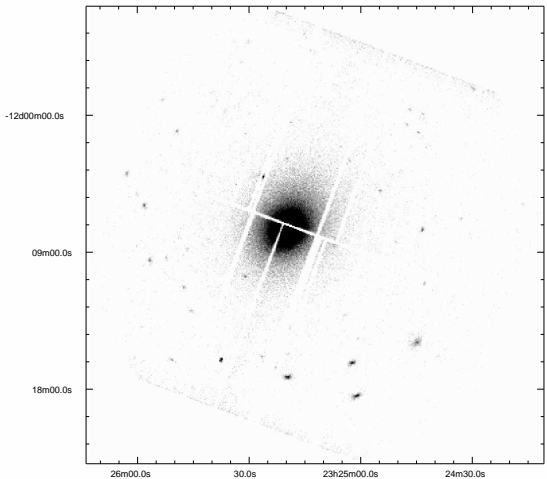}
  \caption{%
    pn count image of Abell~2597, selected with \evsel{FLAG == 0 \&\&
      PATTERN $\leqslant$ 4 \&\& PI in [200:15000]}. The image has
    been GTI filtered and corrected for OOT events as described in
    Section~\ref{sec:data}, but no vignetting or background correction
    has been applied. Image pixels are 4\unit{arcsec} across, roughly
    the size of the PSF core.%
  }
  \label{fig:pnimg}
\end{figure}

The pn image of the cluster is shown in Fig.~\ref{fig:pnimg}. As in
the \rosat{} observation of \citet{sara97}, there are a number of
non-ICM sources in the field of view, though only two of these are
within 6\unit{arcmin} of the cluster centre. The cluster emission
itself displays a relaxed, somewhat elliptical form.

In order to examine the global spectral properties of the cluster, and
to check the agreement between the three EPIC detectors, we extracted
spectra from an annulus of radius 1--4\unit{arcmin} centred on the
cluster emission peak (J2000: RA = 23h25m19.8s, Dec = -12d07m26s; sky
coordinates: $X$ = 23720.5, $Y$ = 23720.5). This radial range was
selected to excise any cooling flow region and so that the background
contribution was insignificant, in order to have the most simple
spectral shape possible. The results of basic single temperature fits
are presented in Table~\ref{tab:spec_1-4arcmin}.

\begin{table}
  \centering
  \caption[]{%
    Global (1--4\unit{arcmin}) spectral properties.
  }
  \begin{tabular}{ccccc}\hline
        &              & A                         & B                      & C                          \\\hline
        & \mysub{N}{H} & 2.49                      & $1.47_{-0.3}^{+0.3}$   & 2.49                       \\
        & $kT$         & $3.55_{-0.05}^{+0.05}$    & $3.69_{-0.07}^{+0.07}$ & $3.55_{-0.05}^{+0.05}$     \\
    M1  & Z            & $0.25_{-0.02}^{+0.02}$    & $0.26_{-0.02}^{+0.02}$ & $0.26_{-0.03}^{+0.02}$     \\
        & z            & 0.0822                    & 0.0822                 & $0.078_{-0.002}^{+0.001}$  \\
        & $\chi^{2}$   & $349.6/343$               & $340.5/342$            & $342.1/342$                \\\hline
        & \mysub{N}{H} & 2.49                      & $1.91_{-0.3}^{+0.3}$   & 2.49                       \\
        & $kT$         & $3.43_{-0.05}^{+0.05}$    & $3.50_{-0.07}^{+0.07}$ & $3.42_{-0.05}^{+0.08}$     \\
    M2  & Z            & $0.22_{-0.02}^{+0.02}$    & $0.23_{-0.02}^{+0.02}$ & $0.22_{-0.02}^{+0.02}$     \\
        & z            & 0.0822                    & 0.0822                 & $0.081_{-0.002}^{+0.002}$  \\
        & $\chi^{2}$   & $297.3/348$               & $294.2/347$            & $296.8/347$                \\\hline
        & \mysub{N}{H} & 2.49                      & $0.01_{-0.01}^{+0.2}$  & 2.49                       \\
        & $kT$         & $3.24_{-0.003}^{+0.030}$  & $3.59_{-0.04}^{+0.04}$ & $3.22_{-0.03}^{+0.02}$     \\
    pn  & Z            & $0.20_{-0.01}^{+0.01}$    & $0.25_{-0.01}^{+0.01}$ & $0.22_{-0.01}^{+0.01}$     \\
        & z            & 0.0822                    & 0.0822                 & $0.077_{-0.001}^{+0.0004}$ \\
        & $\chi^{2}$   & $1321.2/1003$             & $1118.8/1002$          & $1293.6/1002$              \\\hline
        & \mysub{N}{H} & 2.49                      & $0.59_{-0.1}^{+0.1}$   & 2.49                       \\
        & $kT$         & $3.35_{-0.005}^{+0.01}$   & $3.61_{-0.03}^{+0.03}$ & $3.33_{-0.02}^{+0.02}$     \\
    All & Z            & $0.212_{-0.006}^{+0.003}$ & $0.25_{-0.01}^{+0.01}$ & $0.23_{-0.01}^{+0.01}$     \\
        & z            & 0.0822                    & 0.0822                 & $0.079_{-0.001}^{+0.001}$  \\
        & $\chi^{2}$   & $2000.6/1698$             & $1815.8/1697$          & $1968.0/1697$              \\\hline
  \end{tabular}
  \label{tab:spec_1-4arcmin}

  \raggedright
  \mysub{N}{H} is Galactic column in units of $10^{20}\unit{cm^{-2}}$;
  $kT$ temperature in keV; Z metallicity relative to solar. Errors are
  $1\sigma$. Quantities without errors were fixed. Models A--C are
  \phabs $\times$ \xmekal, with: Galactic \mysub{N}{H}; free \mysub{N}{H};
  free $z$. When fitting all three detectors simultaneously, the
  normalizations were untied (though the effects of this were slight
  in this instance).
\end{table}

When a fixed, Galactic absorption is used (model A), the pn has a
temperature somewhat below that of the MOS instruments, and a higher
reduced $\chi^{2}$. If the absorbing column is allowed to vary (model
B), then all three detectors prefer a value significantly below
Galactic, with the pn \mysub{N}{H} being much lower than that of
either MOS. Importantly, however, freeing \mysub{N}{H} results in the
MOS and pn $T$ and $Z$ values coming in to much closer agreement with
each other. The conclusion we may draw from this is that all three
detectors exhibit some form of soft excess (whose origin is unclear)
with respect to a Galactic absorbing column, more so in the pn; but
that when the absorbing column is allowed to be free, the three
detectors agree reasonably well with regards to the other cluster
properties. When all three detectors are analysed simultaneously, the
fits adopt a compromise between the pn and MOS values (there are
roughly equal numbers of counts in the pn and two MOS combined).

Model C has a fixed \mysub{N}{H}, but freely varying redshift $z$. The
three detectors adopt disparate redshifts, with MOS1 and the pn
disagreeing (at the 1$\sigma$ level) with the nominal optical
redshift, 0.0822. Analysis of the RGS data, however, with its higher
spectral resolution (see Section~\ref{sec:rgs_broad-band_fits}),
results in a redshift consistent with the optical value. Comparing the
results of models A and C (which differ only by whether redshift is
free or not), we see that the only effects of freeing $z$ are to
reduce $\chi^{2}$ somewhat -- the results for other parameters such as
$T$ and $Z$ are unchanged. Accordingly, we have left $z$ fixed at the
optical value in all subsequent EPIC fits.

\begin{table*}
  \centering
  \caption[]{%
    Global (0--4\unit{arcmin}) spectral properties.
  }
  \begin{tabular}{cccccccc}\hline
        &              & A                         & B                         & D                         & E                         & F                         & G                        \\\hline
        & \mysub{N}{H} & 2.49                      & $1.49_{-0.2}^{+0.2}$      & 2.49                      & $2.52_{-0.4}^{+0.4}$      & 2.49                      & $2.11_{-0.3}^{+0.3}$     \\
        & $kT$         & $3.35_{-0.03}^{+0.03}$    & $3.47_{-0.04}^{+0.04}$    & $3.46_{-0.03}^{+0.04}$    & $3.45_{-0.03}^{+0.04}$    & $3.56_{-0.05}^{+0.05}$    & $3.56_{-0.05}^{+0.05}$   \\
        & $kT_{2}$     & -                         & -                         & $0.47_{-0.07}^{+0.07}$    & $0.46_{-0.08}^{+0.08}$    & -                         & -                        \\
    M1  & $K_{2} / K$  & -                         & -                         & $0.015_{-0.002}^{+0.002}$ & $0.015_{-0.005}^{+0.005}$ & -                         & -                        \\
        & $\dot{M}$    & -                         & -                         & -                         & -                         & $93_{-15}^{+15}$          & $73_{-19}^{+20}$         \\
        & Z            & $0.35_{-0.01}^{+0.02}$    & $0.37_{-0.02}^{+0.02}$    & $0.39_{-0.01}^{+0.01}$    & $0.39_{-0.02}^{+0.02}$    & $0.37_{-0.02}^{+0.02}$    & $0.37_{-0.02}^{+0.02}$   \\
        & $\chi^{2}$   & $474.9/396$               & $450.0/395$               & $432.0/394$               & $432.1/393$               & $437.4/395$               & $435.4/394$              \\\hline
        & \mysub{N}{H} & 2.49                      & $1.78_{-0.2}^{+0.2}$      & 2.49                      & $2.13_{-0.3}^{+0.3}$      & 2.49                      & $2.56_{-0.3}^{+0.3}$     \\
        & $kT$         & $3.22_{-0.03}^{+0.03}$    & $3.30_{-0.03}^{+0.03}$    & $3.32_{-0.03}^{+0.03}$    & $3.35_{-0.04}^{+0.05}$    & $3.40_{-0.04}^{+0.05}$    & $3.40_{-0.04}^{+0.05}$   \\
        & $kT_{2}$     & -                         & -                         & $0.75_{-0.09}^{+0.1}$     & $0.86_{-0.1}^{+0.2}$      & -                         & -                        \\ 
    M2  & $K_{2} / K$  & -                         & -                         & $0.016_{-0.003}^{+0.003}$ & $0.015_{-0.006}^{+0.006}$ & -                         & -                        \\
        & $\dot{M}$    & -                         & -                         & -                         & -                         & $92_{-16}^{+16}$          & $96_{-22}^{+22}$         \\
        & Z            & $0.30_{-0.01}^{+0.01}$    & $0.31_{-0.01}^{+0.01}$    & $0.32_{-0.02}^{+0.02}$    & $0.32_{-0.02}^{+0.02}$    & $0.31_{-0.01}^{+0.01}$    & $0.31_{-0.01}^{+0.01}$   \\
        & $\chi^{2}$   & $404.3/400$               & $391.9/399$               & $375.5/398$               & $374.0/397$               & $372.0/399$               & $371.9/398$              \\\hline
        & \mysub{N}{H} & 2.49                      & $0.47_{-0.1}^{+0.1}$      & 2.49                      & $0.87_{-0.1}^{+0.1}$      & 2.49                      & $1.11_{-0.1}^{+0.1}$     \\
        & $kT$         & $3.04_{-0.02}^{+0.02}$    & $3.29_{-0.02}^{+0.02}$    & $3.20_{-0.02}^{+0.02}$    & $2.29_{-0.08}^{+0.08}$    & $3.43_{-0.04}^{+0.03}$    & $3.43_{-0.03}^{+0.03}$   \\
        & $kT_{2}$     & -                         & -                         & $0.31_{-0.02}^{+0.02}$    & $6.78_{-0.6}^{+0.5}$      & -                         & -                        \\
    pn  & $K_{2} / K$  & -                         & -                         & $0.035_{-0.003}^{+0.003}$ & $0.51_{-0.07}^{+0.07}$    & -                         & -                        \\
        & $\dot{M}$    & -                         & -                         & -                         & -                         & $159_{-10}^{+8}$          & $86_{-11}^{+11}$         \\
        & Z            & $0.278_{-0.008}^{+0.008}$ & $0.318_{-0.009}^{+0.009}$ & $0.325_{-0.006}^{+0.01}$  & $0.27_{-0.01}^{+0.01}$    & $0.301_{-0.009}^{+0.009}$ & $0.322_{-0.009}^{+0.01}$ \\
        & $\chi^{2}$   & $1781.8/1190$             & $1482.2/1189$             & $1537.3/1188$             & $1307.4/1187$             & $1504.2/1189$             & $1417.2/1188$            \\\hline
  \end{tabular}
  \label{tab:spec_0-4arcmin}

  \raggedright
  Models: A, B $\phabs{} \times \xmekal$; D, E $\phabs{} \times (
  \xmekal + \xmekal )$; F, G $\phabs{} \times (\xmekal + \mkcflow )$.
  The first, second model in each pair have Galactic, free \mysub{N}{H}
  respectively. $K_{2} / K$ relative normalization of second temperature
  component; $\dot{M}$ mass deposition rate
  (\unit{\Msun}\unit{yr^{-1}}); $Z$ metallicity (solar units).

\end{table*}

In Table~\ref{tab:spec_0-4arcmin} are shown the results of fitting the
entire central 4\unit{arcmin}. Now that we include the cooling region,
a wider variety of spectral models become relevant. Models A and B are
the same as in the 1--4\unit{\arcmin} case. The fits in the
0--4\unit{arcmin} region adopt lower temperatures and higher
abundances, so that we may expect to see temperature and abundance
profiles that decrease and increase respectively with decreasing
radius (see Section~\ref{sec:radial_props}).

The results of model B can be compared directly with those of
\citet{alle01}, who found, using \asca: $T = 3.38_{-0.07}^{+0.06}$, $Z
= 0.35_{-0.04}^{+0.03}$, and $\mysub{N}{H} = 0.99_{-0.08}^{+0.07}$ (90
per cent confidence limits).

In models D and E we add a second \xmekal{} component, with an
independent temperature and normalization. For all three detectors,
the fits are significantly improved. For example, for MOS1 with a free
\mysub{N}{H}, the change in fit statistic is $\Delta \chi^{2} = -17.9$
for the removal of two degrees of freedom. According to the F-test,
the significance of such an improvement is ($1 - \sform{3.4e-4}$).
Adhering to strict mathematical formalism, the F-test is not really
valid in cases such as this (testing for the presence of a second
component), where one of the hypotheses lies on the boundary of the
parameter space \citep{prot02}. In practice, however, Monte Carlo
simulations show that it gives acceptable answers \citep{john02}.

The relative normalization of the second temperature component is a
few per cent of that of the bulk hot gas. If the hot and cold phases
are in pressure equilibrium, then $\mysub{n}{C} \mysub{T}{C} =
\mysub{n}{H} \mysub{T}{H}$. The normalization of the \xmekal{} model
varies as $K \propto n^{2} V$. Hence, the relative filling factor of
the cold phase is given by
\[
\mysub{V}{C} / \mysub{V}{H} =
(\mysub{K}{C} / \mysub{K}{H} ) *
(\mysub{n}{H} / \mysub{n}{C})^{2} =
(\mysub{K}{C} / \mysub{K}{H}) *
(\mysub{T}{C} / \mysub{T}{H})^{2};
\]
which applied to the results in Table~\ref{tab:spec_0-4arcmin} for
each detector gives values $\sim$ few $\times$ $10^{-4}$ in this case.

Interestingly, when a second temperature component is available, then
the fitted \mysub{N}{H} for the MOS detectors becomes consistent with
the Galactic value. The pn results, however, for model E are not
sensibly constrained. It is noticeable that the best-fitting
low-temperature components for the two MOS detectors in models D and E
are somewhat discrepant. If (as we will argue later on) there really
is a cooling flow operating down to low temperatures in this system,
then there will be gas present at a continuous range of temperatures,
and it is not obvious what the preferred single low-temperature value
will be in a model with only two temperature components, nor how it
will vary between different instruments. The upper temperatures for
the two MOS instruments also seem to be systematically different,
albeit at a lower level than for the low temperatures. In terms of the
fit, the higher value for \mysub{T}{2} in MOS2 seems to be due to a
slightly broader iron L hump at 1\unit{keV} in this detector.

For completeness, we note that the pn results are stable against
alternative forms of light-curve filtering and background correction.
For example, if we apply the same count-rate cut
(0.45\unit{ct}\unit{s^{-1}}) as used in the blank-sky fields of
\citet{lumb02}, and then employ a double background subtraction
technique (as used for the MOS), then the pn fits are essentially
unchanged.

Using a cooling flow model (in which gas cools from the temperature of
the bulk hot gas), instead of a single temperature, for the second
component, also provides a significant improvement in fit over a
one-component model. The data are not able to distinguish between
these two possibilities with any degree of accuracy. For example, for
MOS1 a cooling flow provides a very slightly worse fit than a single
temperature, whereas for the MOS2 the situation is reversed. The
cooling flow fit does have only one extra free parameter though. When
the absorbing column is allowed to be free, all three detectors
independently fit a mass deposition rate $\sim$
90\unit{\Msun}\unit{yr^{-1}}. An analysis of the Reflection Grating
Spectrometer (RGS) data (Section~\ref{sec:rgs}) produces very similar
mass deposition results.

\begin{figure}
  \centering
  \includegraphics[angle=0,width=0.9\columnwidth]{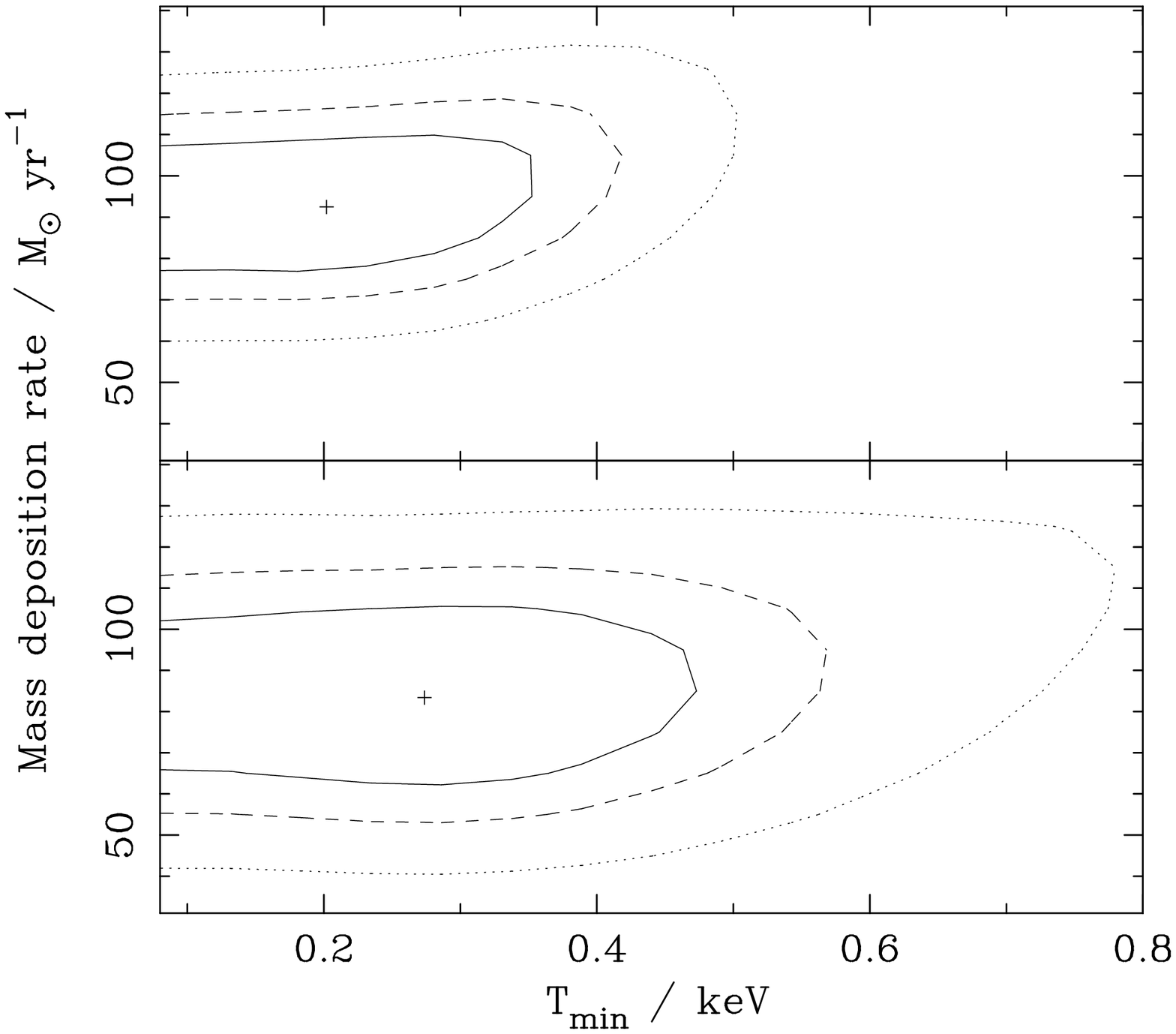}
  \caption{%
    Confidence contours in the $\dot{M}$--\mysub{T}{min} plane for the
    MOS1 + MOS2 cooling flow fits (models F and G in
    Table~\ref{tab:spec_0-4arcmin}, but with \mysub{T}{min} free to
    vary). Upper panel fixed \mysub{N}{H}, lower panel free
    \mysub{N}{H}. The cross shows the best-fit value, and the contours
    are plotted for $\Delta \chi^{2}$ = 2.3, 4.61, 9.21; corresponding
    to the 68, 90, and 99 per cent confidence limits respectively on
    two free parameters. The left-hand boundary is due to the lowest
    usable model temperature, 0.081\unit{keV}.
  }%
  \label{fig:mos_contour}
\end{figure}

In Fig.~\ref{fig:mos_contour} we show the results of allowing the
minimum temperature of the cooling flow component, \mysub{T}{min}, to
vary freely; for both fixed and free \mysub{N}{H} (i.e.\ models F and
G in Table~\ref{tab:spec_0-4arcmin} respectively). When \mysub{T}{min}
is allowed to vary freely, a value somewhat above `zero' (i.e.\ the
model minimum, 0.081\unit{keV}) is preferred; but a zero cut-off
temperature is entirely consistent with the data, within the 1$\sigma$
limits. The associated mass-deposition rate, $\dot{M}$, is little
changed. Freeing \mysub{N}{H} broadens the allowed confidence regions,
but does not greatly affect the results.

\subsection{Comparison with \chandra}\label{sec:chandra_global}

We present a brief comparison of the results of \xmm{} and \chandra{}
observations of Abell 2597 through a re-analysis of the \chandra{}
observation of \citet{mcna01}. We use a conservative flare screening
of the \chandra{} data that results in 7\unit{ks} of good time. The
agreement between the \chandra{} and \xmm{} pointings for the location
of the X-ray emission peak is excellent. The largest complete circle
centred on this point that can be extracted from the ACIS-S3 chip has
a radius of 1\unit{arcmin}. This encompasses most of the cooling
region (see Section~\ref{sec:spec_profiles}). A background spectrum
was extracted from the appropriate standard blank-sky field. The
spectrum was fitted over the 0.6--7.0\unit{keV} range.

\begin{figure}
  \centering
  \includegraphics[angle=0,width=0.9\columnwidth]{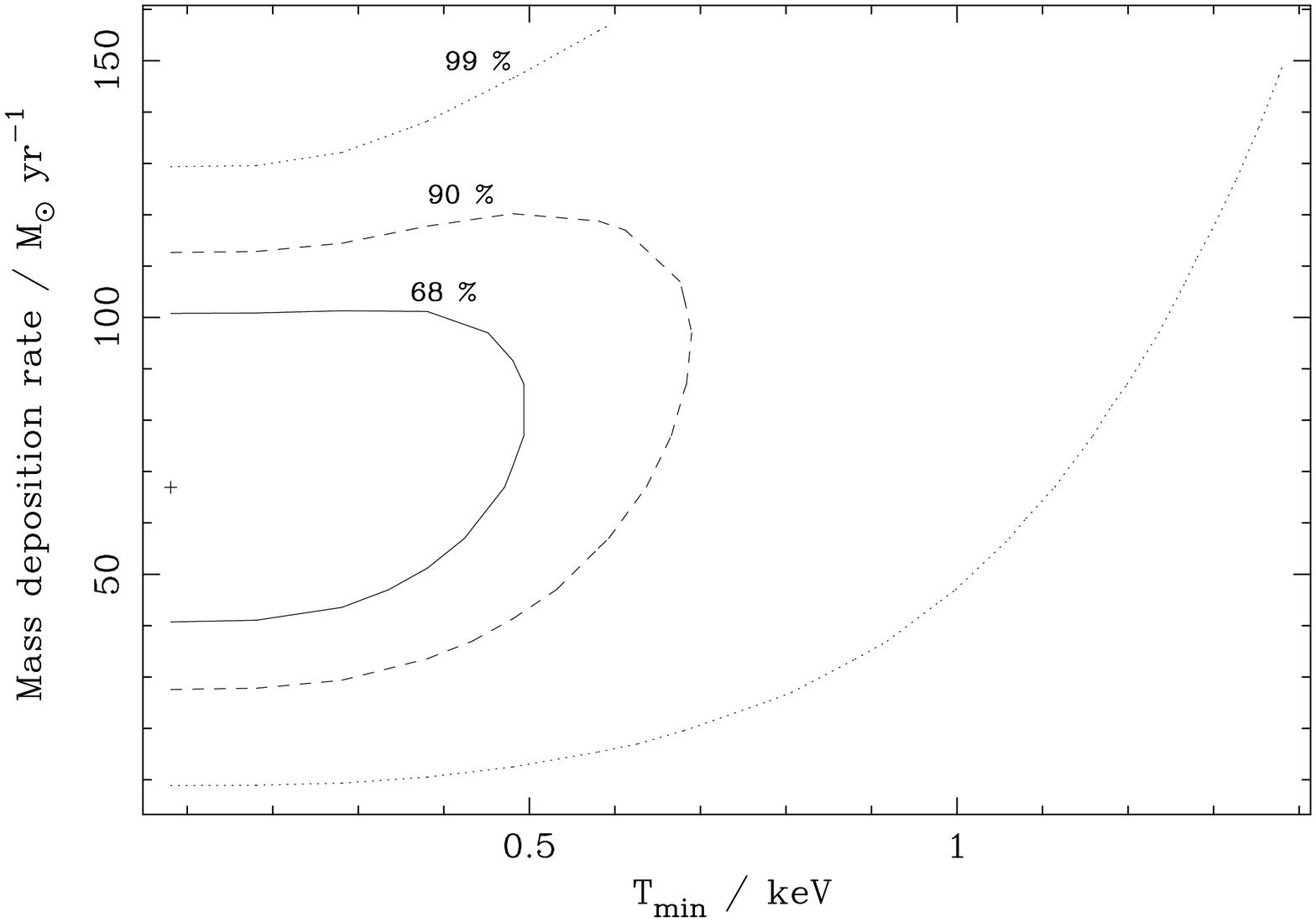}
  \caption{%
    Confidence contours in the $\dot{M}$--\mysub{T}{min} plane for the
    central 1\unit{arcmin} in radius (the cooling region) from the
    \chandra{} ACIS-S3 data, with a Galactic \mysub{N}{H}. Plot
    details are as per Fig~\ref{fig:mos_contour}.
  }%
  \label{fig:chandra_contour}
\end{figure}

Fig.~\ref{fig:chandra_contour} is the \chandra{} counterpart of the
upper panel (i.e.\ for a fixed \mysub{N}{H}) of
Fig.~\ref{fig:mos_contour} for the MOS. We were unable to obtain
sensible constraints when the absorbing column was allowed to vary
freely. The preferred low-temperature cut-off for the cooling flow
component, though not tightly constrained, is as low as possible. The
associated mass deposition rate is $\sim$
70\unit{\Msun}\unit{yr^{-1}}, entirely consistent with the \xmm{} EPIC
results for the central 4\unit{arcmin}. These results are also in
agreement with those from the RGS, as shown in
Fig.~\ref{fig:rgs_contour}. Fig~\ref{fig:chandra_contour} may be
contrasted with the more common \chandra{} result in such cases, where
a much higher \mysub{T}{min} is preferred \citep[e.g. in
Abell~3581;][fig.~10]{john05}.

\section{Radial properties}\label{sec:radial_props}

Given the symmetric nature of the emission visible in
Fig.~\ref{fig:pnimg}, we have performed analyses using circular annuli
centred on the cluster emission peak.

\subsection{Surface brightness profile}\label{sec:sb_profile}

\begin{figure}
  \centering
  \includegraphics[angle=0,width=0.9\columnwidth]{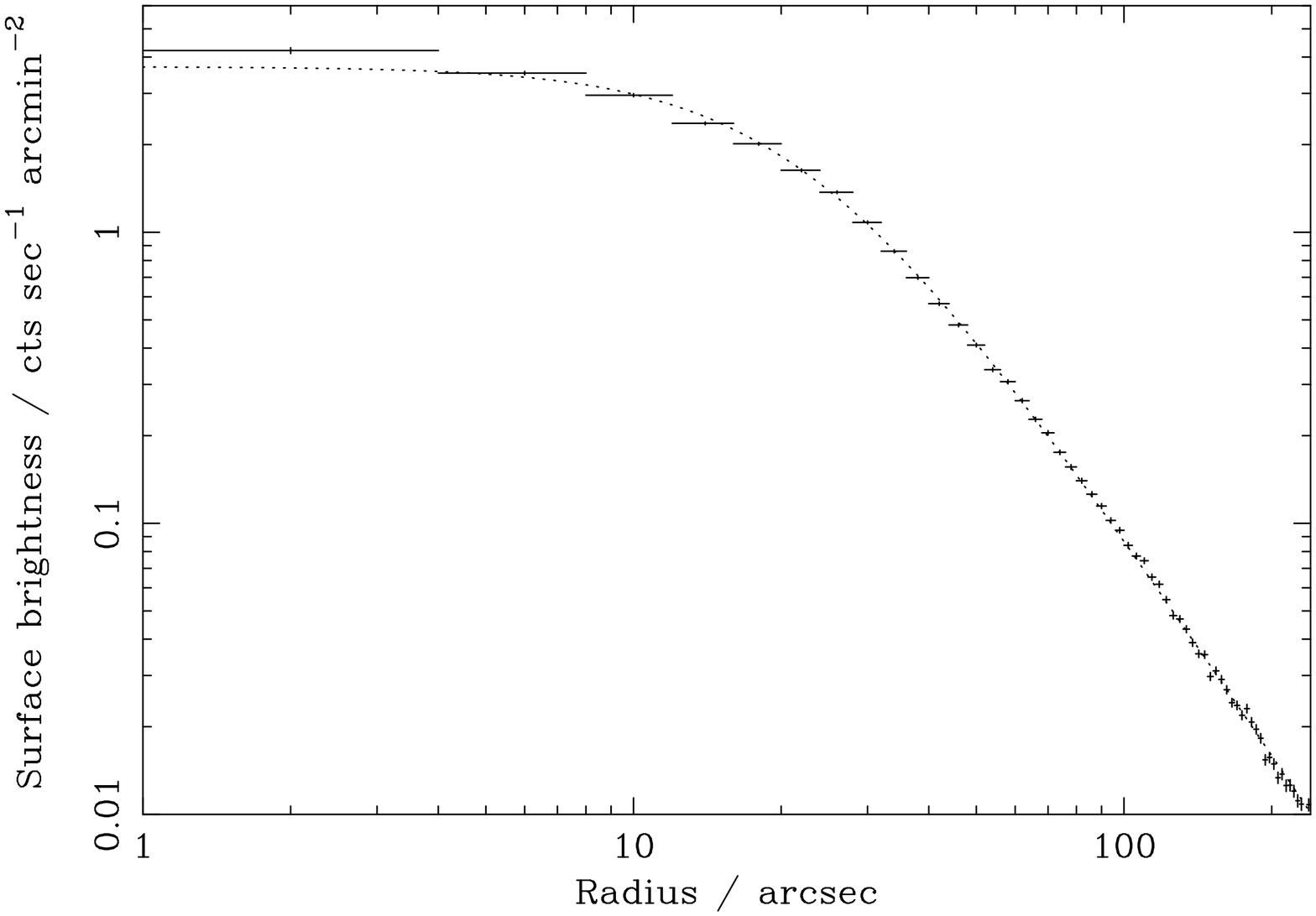}
  \caption{%
    Surface brightness profile of the central 240\unit{arcsec} in
    radius, 0.5--2.0\unit{keV} band, from summed MOS1 and MOS2
    profiles. The dotted line shows a $\beta$ profile fit to the data
    -- see text for details. The profile has been corrected for
    background and vignetting.
    }%
  \label{fig:sb_mos}
\end{figure}

Fig.~\ref{fig:sb_mos} shows the surface brightness profile of the
central 4\unit{arcmin} in radius of the cluster in the
0.5--2.0\unit{keV} band, from the sum of the MOS1 and MOS2 data. The
profile was extracted from 1\unit{arcsec} pixel images and
vignetting-corrected exposure maps, using circular annuli centred on
the emission peak. A background was subtracted using the appropriate
scaled blank-field data sets. The width of each annulus was such that
the significance of the background-subtracted counts it contained was
at least $5\sigma$. Point sources were excluded.

Also shown is the result of a $\beta$ profile fit to the surface
brightness, $\Sigma(r) = \Sigma_{0} [ 1 + (r/\mysub{r}{c})^{2} ]^{-3
  \beta + 0.5}$; with $\beta = 0.585_{-0.001}^{+0.002}$, $\mysub{r}{c}
= 23.0_{-0.02}^{+0.02}\unit{arcsec}$.

The surface brightness profile is affected by the \xmm{} PSF, which
can itself be well described by a $\beta$ model \citep{ghiz01,ghiz02},
with a core radius of around 5\unit{arcsec}. When an intrinsic
brightness profile $a(\vec{r})$ is modified by an instrumental PSF
$b(\vec{r})$, the observed profile is the convolution of the two,
\begin{equation}
  c(\vec{r}) = \int \diff^{2} \vec{u} \; a(\vec{u}) b(\vec{r} - \vec{u})
   = 2 \pi \int_{0}^{\infty} \diff k \; k J_{0} (k r)
   \tilde{a}(k) \tilde{b}(k),
\end{equation}
where the second form is for the case of radial symmetry and makes use
of the convolution theorem. $\tilde{a}(k)$ in this expression is the
radially symmetric Fourier transform \citep{birk94}, or Hankel
transform of order zero,
\begin{equation}
\tilde{f}(k) = \int_{0}^{\infty} \diff r \; J_{0} (k r) r f(r),
\end{equation}
with $J_{0}$ a Bessel function of the first kind. Evaluating the
effect of instrument PSF on a surface brightness profile therefore
reduces to performing three Hankel transforms. The general $\beta$
function has an analytic Hankel transform \citep[][equation
6.565.4]{grad00}, but the specific case of $\beta = 2/3$, which is a
good approximation for both the \xmm{} PSF and many cluster brightness
profiles, is especially simple:
\begin{eqnarray}
  f(r) = (a^{2} + r^{2})^{-3/2} & \Leftrightarrow &
  \tilde{f}(k) = \frac{\mathrm{e}^{-a k}}{a}
\end{eqnarray}
As a result, if
\begin{eqnarray}
  S(r) = S_{0} [1 + (r/\mysub{r}{s})^{2}]^{-3/2} & \mathrm{and} &
  P(r) = P_{0} [1 + (r/\mysub{r}{p})^{2}]^{-3/2}
\end{eqnarray}
describe the source and PSF profiles respectively, with the latter
normalized so that $P_{0} = (2 \pi \mysub{r}{p})^{-1}$, then it is
straightforward to show that the convolution $c(r)$ is given by
\begin{equation}
  c(r) = S_{0}
  \Big( \frac{\mysub{r}{s}}{\mysub{r}{s} + \mysub{r}{p}} \Big)^{2}
  \left[ 1 +
    \Big( \frac{r}{\mysub{r}{s} +  \mysub{r}{p}} \Big)^{2}
  \right]^{-3/2}.
\end{equation}
This is nothing more than another $\beta$ profile, with a reduced
central amplitude $S'_{0}$ and increased core radius $\mysub{r'}{s}$:
\begin{eqnarray}
  S'_{0} = S_{0}
  \Big( 1 + \frac{\mysub{r}{p}}{\mysub{r}{s}} \Big)^{-2}
  \sim S_{0} \Big( 1 - 2 \frac{\mysub{r}{p}}{\mysub{r}{s}} \Big)
  & \mathrm{and} &
  \mysub{r'}{s} = \mysub{r}{s} + \mysub{r}{p}.
\end{eqnarray}

The simplest PSF correction we can make to the surface brightness
profile is therefore to subtract the radius of the PSF,
5\unit{arcsec}, from the core fitted to the convolved profile,
23\unit{arcsec}, resulting in an estimate of 18\unit{arcsec} for the
intrinsic surface brightness core radius.

\subsection{Spectral profiles}\label{sec:spec_profiles}

In order to obtain reliable spectral fits, annular radii were chosen
so as to encompass slightly more than 10000 (net, i.e.\
background-subtracted) counts in each MOS, and thus over 20000 counts
in the pn. It was found that annuli with fewer counts (e.g.\ 7500 per
MOS) tended to give poor (unphysical) results in the deprojection and
PSF-correction analyses. In such cases, the lower signal-to-noise
allows physically unrealistic solutions (in which fitted quantities
tend to oscillate about the smoother profiles obtained from annuli
with more counts) to become mathematically valid. Both correction
algorithms (see e.g.\ \citealt{john05} for testing of the deprojection
process) involve redistribution of counts between annuli, potentially
(depending on the details of the brightness profile) leaving small net
numbers of counts in any one annulus. In practice it is found to be
beneficial not to let the uncorrected count numbers fall too low. The
outermost annulus was truncated at a radius of 5\unit{arcmin} (due to
the increasing relative background) and so contains somewhat fewer
counts. By the time this last annulus is reached, the background is
contributing about 30 per cent of the total counts.

\subsubsection{Projected profiles}

\begin{figure}
  \centering
  \includegraphics[angle=0,width=0.9\columnwidth]{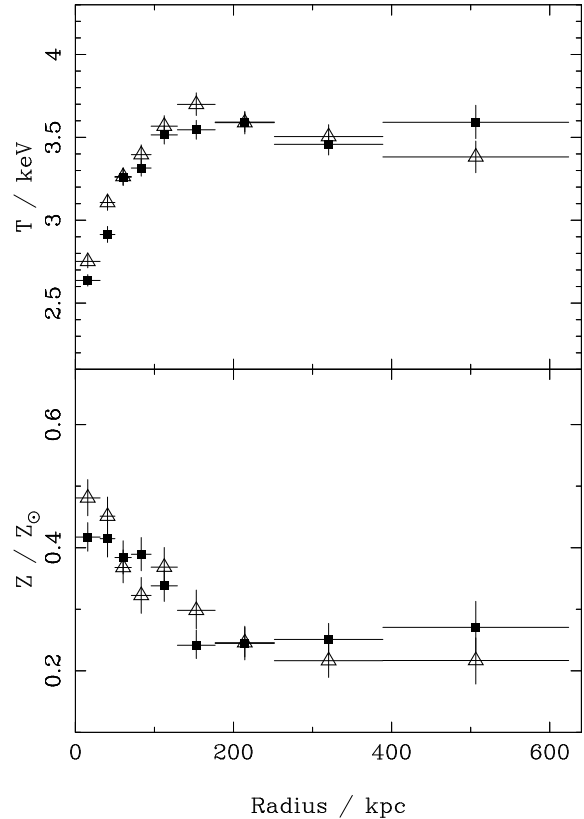}
  \caption{%
    Projected temperature (upper panel) and metallicity (lower panel)
    profiles. The open triangles are simultaneous MOS1 + MOS2 fits,
    the filled squares are the pn. The model was $\phabs \times
    \xmekal$ with a common, free absorbing column \mysub{N}{H} shared
    by all annuli. The \mysub{N}{H} values were (in units of
    $10^{20}$\unit{cm^{-2}}): $1.52_{-0.08}^{+0.08}$ (MOS);
    $0.19_{-0.06}^{+0.06}$ (pn). With \mysub{N}{H} free to vary
    between annuli, the T and Z profiles are not much changed. The MOS
    \mysub{N}{H} shows no significant variation with radius; whereas
    that of the pn is consistent with the MOS in the central two bins,
    but much lower outside this region.
    }%
  \label{fig:TZ_proj}
\end{figure}

Fig.~\ref{fig:TZ_proj} shows the results of fitting a
single-temperature \mekal{} \citep{mewe95} model to each spectral
annulus, with an uniform (but free to vary) absorbing column applied
to all annuli. The central temperature is a little over 2.5\unit{keV}
(a function of the size of the central annulus). Over the central
100\unit{kpc} or so the temperature increases, reaching a maximum of
$\sim 3.6$\unit{keV}. Outside the central cooling region, there is
perhaps evidence for a gradual decline in temperature beyond $\sim
150$\unit{kpc} or so. The metallicity exhibits a fairly smooth decline
with increasing radius in the central cooling region, dropping from
around 0.5\unit{\Zsun} at the centre to 0.2\unit{\Zsun} at the
outside.

Results are shown for the simultaneous fit of both MOS instruments,
and also for the pn. The two instruments agree reasonably well in
terms of $T$ and $Z$, but have very different preferences for
\mysub{N}{H}.

\begin{figure}
  \centering
  \includegraphics[angle=0,width=0.9\columnwidth]{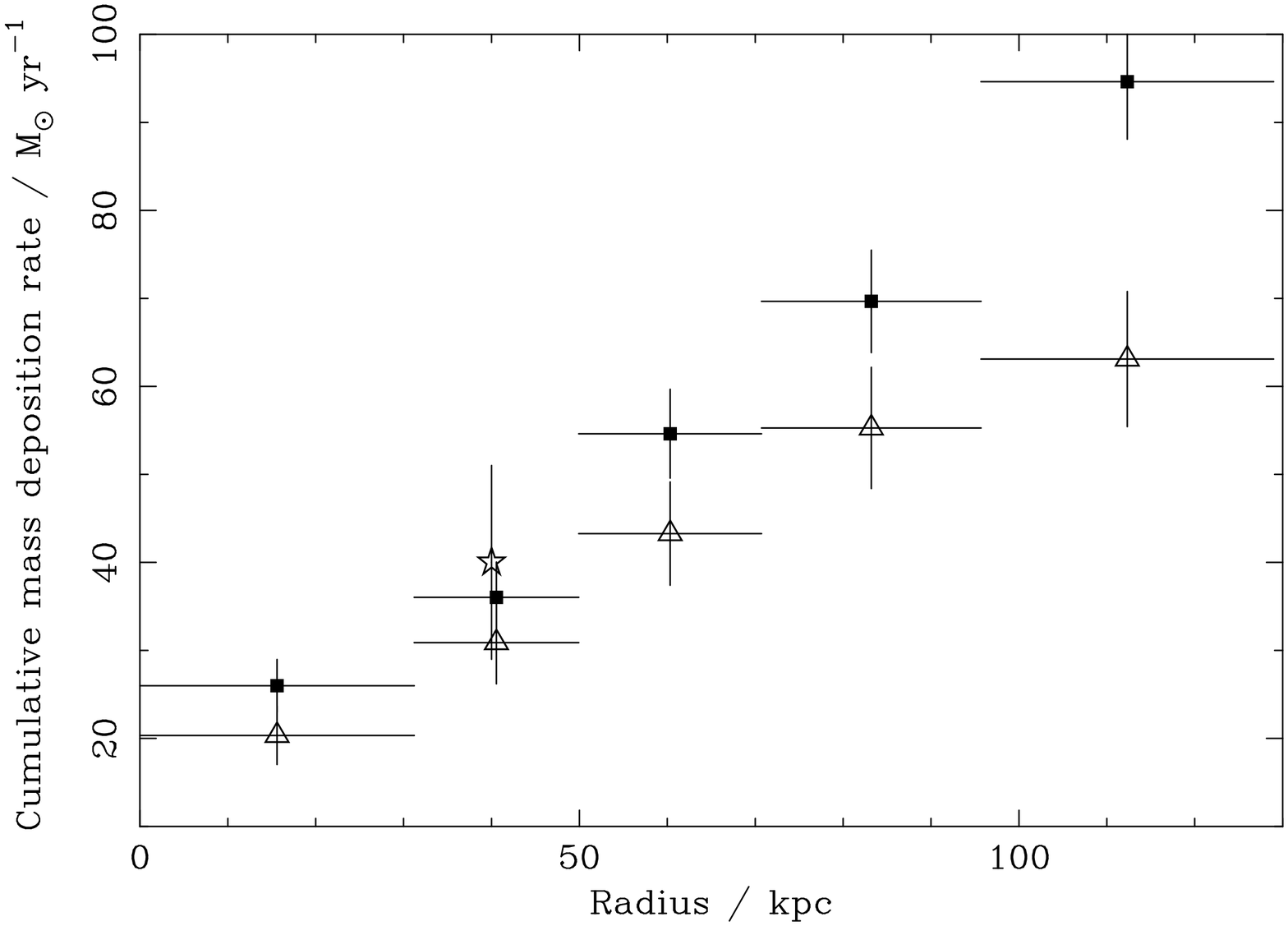}
  \caption{%
    Cumulative projected mass deposition rate, $\dot{M}(<r)$, for the
    cooling flow region. The open triangles are simultaneous MOS1 +
    MOS2 fits, the filled squares are the pn. The model was $\phabs
    \times (\xmekal +\mkcflow$) with a fixed Galactic absorption. The
    star shows the \fuse{} \ion{O}{6} data point of \citet{oege01},
    with an error bar appropriate for the accuracy of the \fuse{} flux
    measurement, $(1.3 \pm 0.35) \times
    10^{-15}\unit{erg}\unit{cm^{-2}}\unit{s^{-1}}$.
  }%
    \label{fig:mdot_cumu}
\end{figure}

In Fig.~\ref{fig:mdot_cumu}, we show the radial dependence of the
cumulative mass deposition rate, $\dot{M}(<r)$, within the cooling
flow radius (130\unit{kpc}; see Section~\ref{sec:deproj}). There is a
discrepancy between the MOS and pn detectors, with the former
preferring smaller mass deposition rates. Owing to the way this figure
was constructed (fitting each annulus independently with a single
temperature component and a cooling flow component, then summing the
$\dot{M}$ values so obtained from the centre outwards), the
discrepancy between the two instruments builds with increasing radius.
Also shown is the $\dot{M}$ value inferred by \citet{oege01} from
their \fuse{} data; namely 40\unit{\Msun}\unit{yr^{-1}} within a
radius of 40\unit{kpc}. The figure illustrates that there is
reasonable agreement between the X-ray and UV mass deposition rates
within a radius of 40\unit{kpc}.

\subsubsection{Deprojected profiles}\label{sec:deproj}

\begin{figure}
  \centering
  \includegraphics[angle=0,width=0.9\columnwidth]{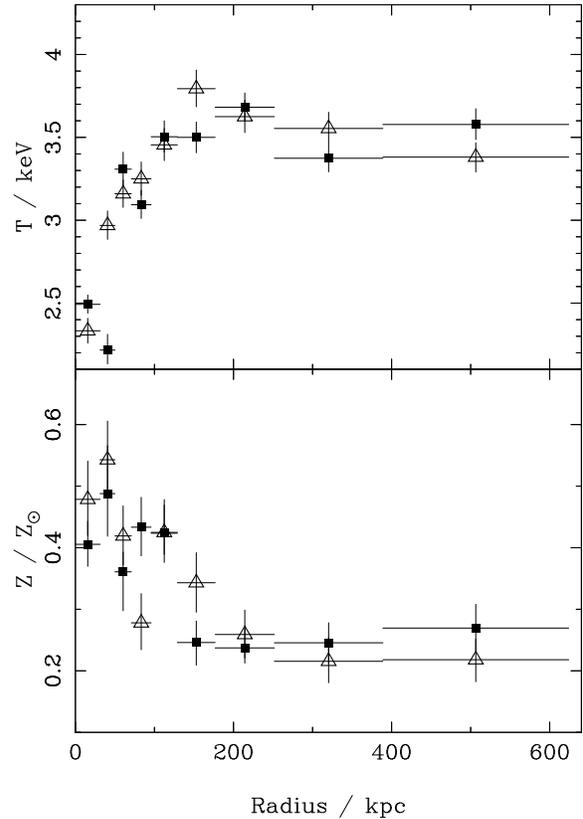}
  \caption{%
    Deprojected temperature (upper panel) and metallicity (lower
    panel) profiles. The open triangles are simultaneous MOS1 + MOS2
    fits, the filled squares are the pn (as per
    Fig.~\ref{fig:TZ_proj}). The model was $\projct \times \phabs
    \times \xmekal$ with a common, free absorbing column \mysub{N}{H}
    shared by all annuli. The \mysub{N}{H} values were (in units of
    $10^{20}$\unit{cm^{-2}}): $1.53_{-0.08}^{+0.08}$ (MOS);
    $0.23_{-0.06}^{+0.06}$ (pn); little changed from the projected
    results.
    }%
  \label{fig:TZ_deproj}
\end{figure}

The results of Fig.~\ref{fig:TZ_proj} are subject to projection
effects, in which the spectral properties at any point in the cluster
are the emission-weighted superposition of radiation originating at
all points along the line of sight through the cluster. In
Fig.~\ref{fig:TZ_deproj} we show the results of a deprojection of the
MOS and pn spectra using the \xspec{} \projct{} model. Under the
assumption of intrinsic spherical (more generally, ellipsoidal) shells
of emission, this model calculates the geometric weighting factors
according to which emission is redistributed amongst the projected
annuli.

As is to be expected, the temperature in the outer regions where the
profile is fairly flat is seen to be relatively little influenced by
projection. Deprojection recovers a lower central temperature than
before, since in the projected fits the spectrum of the central
annulus is contaminated by overlying hotter emission. There is also
some evidence for a higher central abundance in the deprojected
metallicity profile.

The deprojected pn temperature profile shows some signs of instability
in the centre, in which neighbouring bins `oscillate' with high and
low temperatures. The interpolated values, though, agree well with the
MOS results. Recently, \citet{john05} have shown that the \projct{}
approach works well at reproducing various simple synthetic cluster
temperature and density structures. Possibly the slight instability of
the pn fit reflects a more complex under-lying temperature
distribution (e.g.\ more than one phase); but more likely it is simply
a quirk of geometry that renders certain physically unrealistic
solutions mathematically plausible. Freezing the abundances at the
projected value makes little difference, but the effect can be removed
by using larger annuli (although the pn annuli shown already contain
similar numbers of counts to the combined MOS annuli), but we retain
the same annuli here for both instruments for illustration.

\begin{figure}
  \centering
  \includegraphics[angle=0,width=0.9\columnwidth]{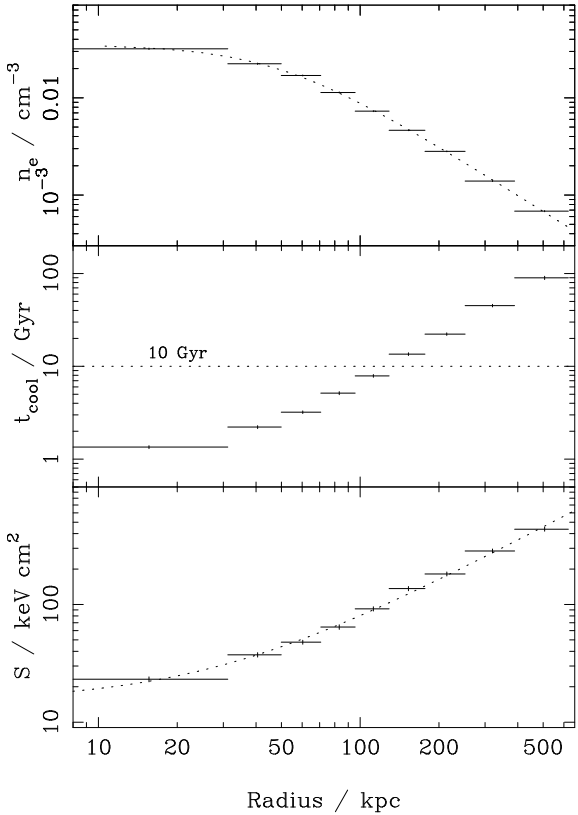}
  \caption{%
    Derived quantities from the deprojected profiles of
    Fig.~\ref{fig:TZ_deproj}: upper panel electron density; middle
    panel approximate isobaric cooling time; lower panel
    entropy. The dotted lines show the fits of some simple models to
    the data: a beta profile for the density; and a constant plus
    power-law for the entropy. See text for details.
  }%
  \label{fig:ne_tcool_S_deproj}
\end{figure}

In Fig.~\ref{fig:ne_tcool_S_deproj} we show various quantities derived
from the deprojected spectral fits. The electron density \mysub{n}{e}
is obtained from the \xmekal{} normalization, and scales as
$H_{0}^{1/2}$. The cooling time (scaling as $H_{0}^{-1/2}$) is an
approximate isobaric one, calculated as the time taken for the gas to
radiate its enthalpy using the instantaneous cooling rate at any
temperature,
\begin{equation}
  \mysub{t}{cool} \approx \frac{H}{\mysub{n}{e}\mysub{n}{H}\Lambda(T)}
  = \frac{\gamma}{\gamma - 1}
  \frac{kT}{\mu \mysub{X}{H} \mysub{n}{e} \Lambda(T)},
\end{equation}
where: $\gamma = 5/3$ is the adiabatic index; $\mu \approx 0.61$ (for
a fully-ionized 0.3\unit{\Zsun} plasma) is the molecular weight;
$\mysub{X}{H} \approx 0.71$ is the hydrogen mass fraction; and
$\Lambda(T)$ is the cooling function. The `entropy' is the standard
astronomer's entropy, $S = T \mysub{n}{e}^{-2/3}$, and scales as
$H_{0}^{-1/3}$.

The cooling time is less than 10\unit{Gyr} inside a radius of
130\unit{kpc} (in perfect agreement with the \rosat{} result of
\citealt{sara95}), and less than 5\unit{Gyr} inside 85\unit{kpc}.

Also shown in Fig.~\ref{fig:ne_tcool_S_deproj} are the results of
simple model fits to the density and entropy profiles. The density
profile is represented by a $\beta$ model, $n(r) = n_{0} [ 1 +
(r/\mysub{r}{c})^{2} ]^{-1.5 \beta}$; with $\beta =
0.57_{-0.01}^{+0.01}$, $n_{0} =
0.0354_{-0.0006}^{+0.0006}\unit{cm^{-3}}$, $\mysub{r}{c} =
49.0_{-0.9}^{+0.9}\unit{kpc}$. These results for the density profile
are in reasonable agreement with those obtained from fitting the
surface brightness profile (Section~\ref{sec:sb_profile}). The absence
of a significant excess in the deprojected density in the last annulus
indicates that we reach a sufficiently large radius that projection
from more distant material is not significant \citep[compare with the
\chandra{} deprojection results of][]{john05}. The entropy profile is
parametrized as a constant plus power-law, $S(r) = S_{0} + k
r^{\eta}$; with $S_{0} = 15.1_{-1.5}^{+1.5}\unit{keV}\unit{cm^{2}}$,
$k = 0.28_{-0.05}^{+0.05}\unit{keV}\unit{cm^{2}}\unit{kpc^{-1.19}}$,
$\eta = 1.19_{-0.04}^{+0.03}$. The fit with the $S_{0}$ term is
substantially better than that without, but bear in mind that the PSF
has not been corrected for here. A logarithmic slope for the entropy
profile of $\approx 1.1$ is produced in simulations involving
gravitational collapse and shock heating \citep{tozz01}.

The virial radius becomes a quantity of interest at this point. It is
known that the calibrations of the scaling relations obtained from
simulations have normalizations quite different to those obtained from
observation. \citet{alle01b} studied the cluster virial scaling
relations using \chandra, but do not quote an explicit
radius--temperature relation. Using their data for an SCDM cosmology,
we calibrate the $r$--$T$ relation as $r_{2500} = 0.31\unit{Mpc}\;
h_{50}^{-1} (1 + z)^{-3/2} (T_{2500} / \unit{keV})^{0.5}$. The
conversion from an overdensity of 2500 to one of 200 (appropriate for
the virial radius) depends upon the form assumed for the density
profile. Taking an NFW profile with a typical cluster concentration $c
= 5$, we find that $r_{200} \approx 2.75 r_{2500}$. Using a
temperature of 3.5\unit{keV}, we therefore estimate a virial radius of
1.4\unit{Mpc} for Abell~2597. The radius 0.1\mysub{R}{V} therefore
lies within the sixth annulus from the centre (a bin whose mid-point
is at 150\unit{kpc}). The measured entropy in this bin is
$S_{0.1\mysub{R}{V}} = (137 \pm 6)
h_{50}^{-1/3}$\unit{keV}\unit{cm^{2}}. This is a relatively low value
\citep[see e.g.\ fig.~4 of][]{ponm03}, but recall PSF effects are
still present in this result.

\subsubsection{PSF-corrected profiles}\label{sec:xmmpsf}

The broadening effect of the \xmm{} PSF acts to redistribute counts
between spectral annuli in a manner conceptually similar to that of
projection. Consequently, it can be corrected for in the same way,
namely by using an \xspec{} mixing model, \xmmpsf.

The most recent version available at the time of writing was unable to
fit observations from different instruments simultaneously, so we
present here only the results for MOS1, in order to illustrate the
effects of PSF on the spectral fits. We have been unable to achieve
physically realistic solutions when using the \xmmpsf{} and \projct{}
models in combination, to correct for both projection and PSF at the
same time. The exception is when annuli that are physically very large
are employed; but in such cases the PSF redistribution becomes less
significant anyway, and we also obtain very limited resolution of the
central cooling region, which is the main area of interest here.

\begin{figure}
  \centering
  \includegraphics[angle=0,width=0.9\columnwidth]{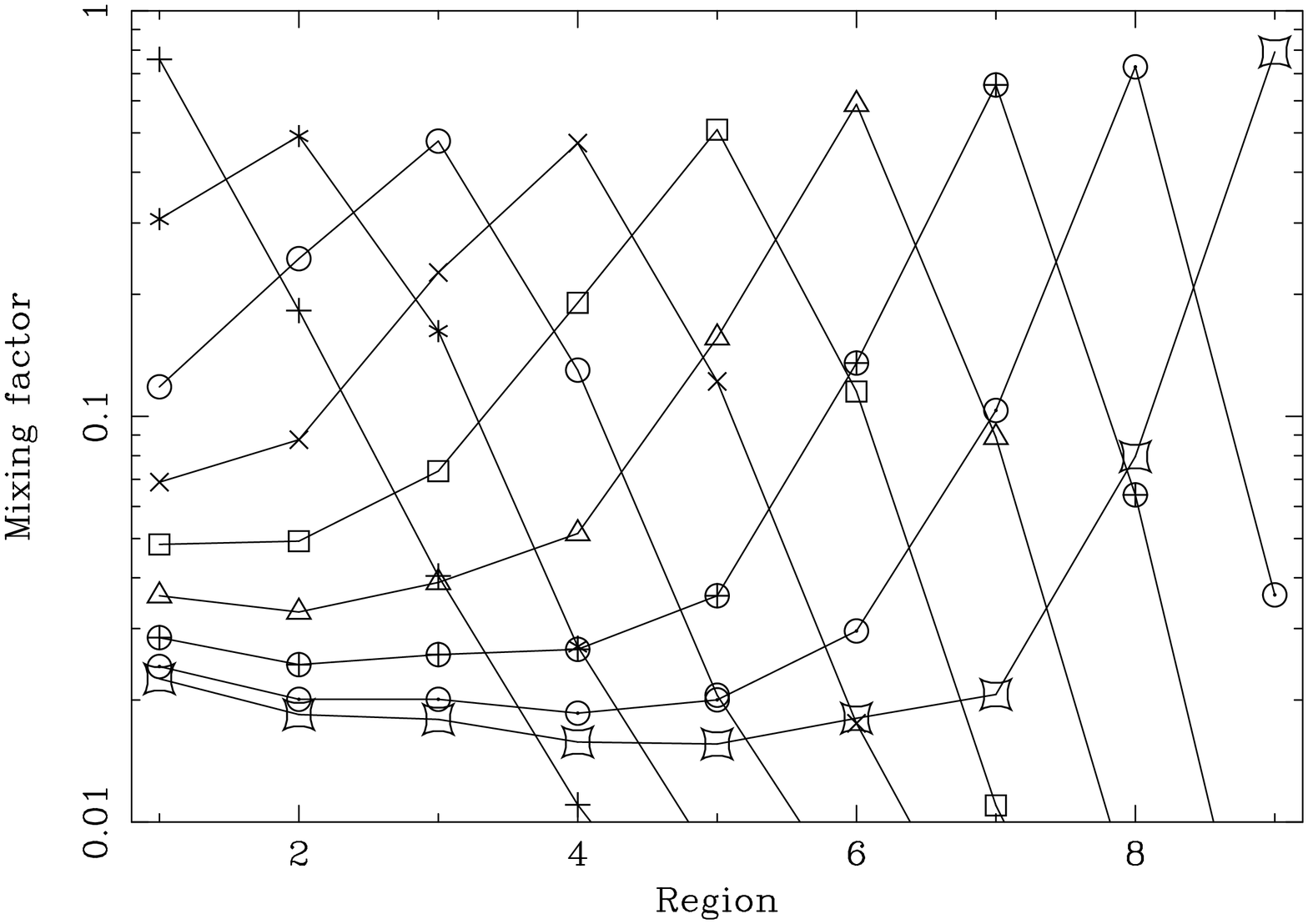}
  \caption{%
    \xmmpsf{} mixing factors calculated at 1.0\unit{keV}. Each curve
    shows, for one of the annular regions, the fraction of the flux in
    that region originating \textit{from} each of the other regions.
    Only contributions greater than 1 per cent are shown. The peak
    flux is always from the region itself. The actual mixing factors
    used internally by the model relate to the fraction of the flux
    from each region going \textit{to} other regions. The factors
    shown here are calculated from these using the model
    normalizations fitted to each annulus. By construction, the points
    with the same x-coordinate sum to 1.0.
  }%
  \label{fig:xmmpsf_fact}
\end{figure}

For a given set of spectral annuli, and a given brightness profile,
the \xmmpsf{} model calculates the mixing factors through which flux
from any given annulus is redistributed to all the others. The factors
are calculated at ten energies in the range 0.5--6.0\unit{keV},
although the energy dependence of the \xmm{} PSF is relatively small.
The input brightness profile can be specified using either an image,
or a model parametrization. We find that there is no significant
difference between using a MOS1 (say) image or the PSF-corrected
$\beta$ profile from Section~\ref{sec:sb_profile} to provide the
brightness information. By way of example, we show the mixing factors
for our set of annuli at 1.0\unit{keV} in Fig.~\ref{fig:xmmpsf_fact}.
At the ten per cent level, each annulus is only affected by its
immediate neighbours. Nevertheless, the mixing is potentially quite
significant, since for many regions only 50 per cent of their flux
originates from that same region.

\begin{figure}
  \centering
  \includegraphics[angle=0,width=0.9\columnwidth]{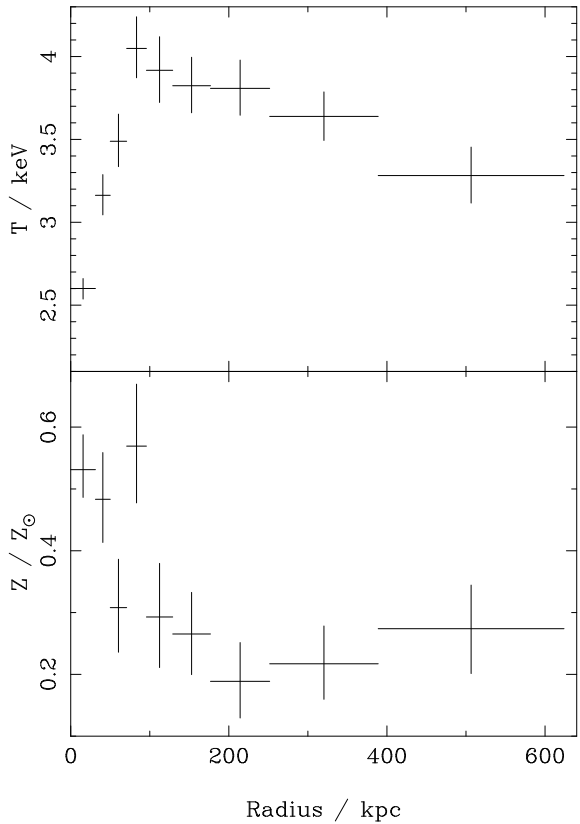}
  \caption{%
    PSF-corrected MOS1 temperature (upper panel) and metallicity
    (lower panel) profiles. The model was $\xmmpsf \times \phabs
    \times \xmekal$ with a common, free absorbing column shared by all
    annuli: $\mysub{N}{H} = 1.40_{-0.1}^{+0.1} 10^{20}\unit{cm^{-2}}$.
  }%
  \label{fig:TZ_xmmpsf}
\end{figure}

In Fig.~\ref{fig:TZ_xmmpsf}, we show the $T$ and $Z$ profiles obtained
from a PSF-corrected fit of the MOS1 data. Compared to the uncorrected
profiles (Fig.~\ref{fig:TZ_proj}), a steeper central temperature
gradient is recovered, as well as a larger central abundance peak.

\begin{figure}
  \centering
  \includegraphics[angle=0,width=0.9\columnwidth]{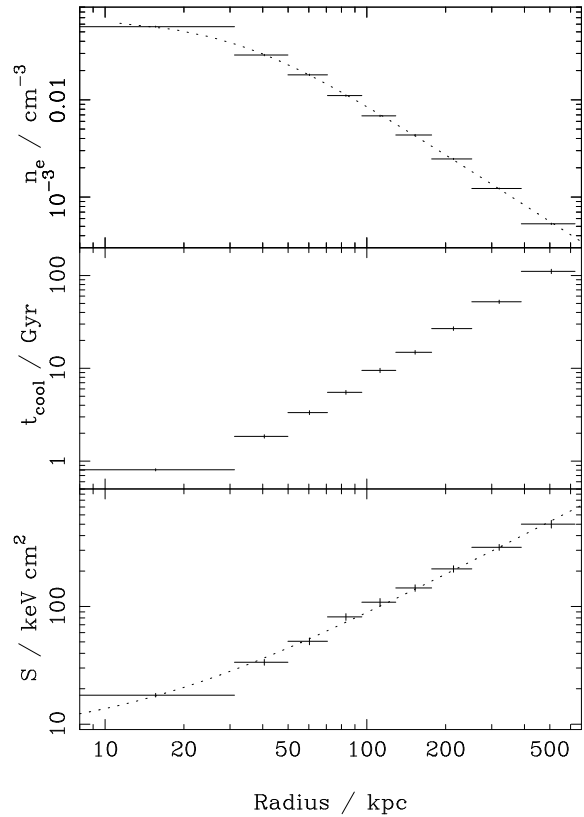}
  \caption{%
    Derived quantities from the MOS1 PSF-corrected profiles of
    Fig.~\ref{fig:TZ_xmmpsf}. Details as per
    Fig.~\ref{fig:ne_tcool_S_deproj}.
  }%
  \label{fig:ne_tcool_S_xmmpsf}
\end{figure}

Fig.~\ref{fig:ne_tcool_S_xmmpsf} displays the density, cooling time,
and entropy derived from the PSF-corrected profiles. Compared to the
results of Fig.~\ref{fig:ne_tcool_S_deproj}, a higher central density,
and a lower central cooling time and entropy are obtained. Also shown
is a $\beta$ model fit for the density, with: $\beta =
0.579_{-0.004}^{+0.004}$, $n_{0} =
0.068_{-0.001}^{+0.001}\unit{cm^{-3}}$, $\mysub{r}{c} =
31.6_{-0.6}^{+0.6}\unit{kpc}$. The PSF corrected core radius of
15\unit{arcsec} is comparable to the estimate of 18\unit{arcsec} made
in Section~\ref{sec:sb_profile} using the surface brightness profile.

The entropy profile is well-fit -- reduced $\chi^{2} = 7.49 / (9~-~3)$
-- by a constant plus power-law model of the same form used in
Section~\ref{sec:deproj}, with: $S_{0} =
7.9_{-1.5}^{+1.3}\unit{keV}\unit{cm^{2}}$, $k =
0.40_{-0.08}^{+0.11}\unit{keV}\unit{cm^{2}}\unit{kpc^{-1.15}}$, $\eta
= 1.15_{-0.04}^{+0.05}$.

\subsection{Comparison with \chandra{} temperature profile}
\label{sec:chandra_profile}

As per Section~\ref{sec:chandra_global}, we have compared the \xmm{}
and \chandra{} results. \chandra{} spectra were extracted from the
same annuli as used in the \xmm{} analysis. Restricting ourselves to
those annuli that lie entirely on the ACIS-S3 chip, we obtain five
spectral regions from the \chandra{} data. Background spectra were
produced as previously described (though background should not be
significant in this region).

\begin{figure}
  \centering
  \includegraphics[angle=0,width=0.9\columnwidth]{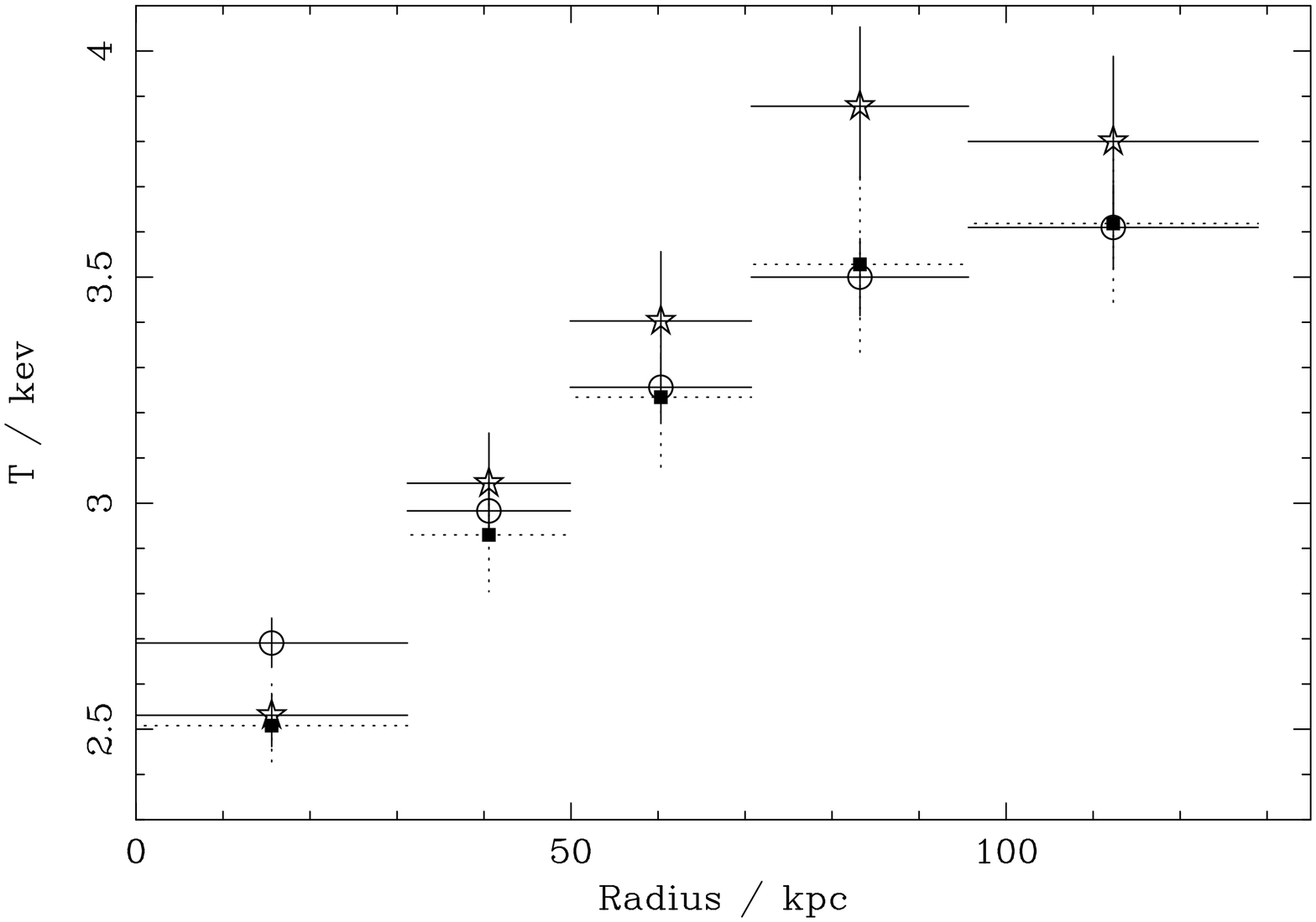}
  \caption{%
    Comparison of MOS1 and \chandra{} ACIS-S3 temperature profiles.
    Open symbols MOS1: circles non-PSF corrected; stars PSF corrected.
    Filled squares \chandra. The model in each case was $\phabs \times
    \xmekal$ with a fixed Galactic \mysub{N}{H}. The PSF correction
    was achieved with the addition of an \xmmpsf{} component (see text
    for details).
    }%
  \label{fig:T_chandra_mos}
\end{figure}

In Fig.~\ref{fig:T_chandra_mos} we compare the temperature profiles
obtained from MOS1 and \chandra, using a simple $\phabs \times
\xmekal$ model with a fixed Galactic absorption. The agreement between
the MOS1 and ACIS temperatures is good everywhere except in the
central bin, where the ACIS temperature is significantly lower than
that of the MOS. Correcting for the \xmm{} PSF using the \xmmpsf{}
model in the previously described manner lowers the central MOS1
temperature so that it agrees very well with that of \chandra. The
agreement of the other MOS1 points with those of ACIS is worsened
though, as the PSF correction raises all but the central MOS1
temperatures.

The \xmm/\chandra{} cross-calibration has recently been examined by
\citet{kirs04}.

\section{Reflection Grating Spectrometer}\label{sec:rgs}

Around half of the light in each telescope feeding the MOS detectors
is diverted to a Reflection Grating Spectrometer
\citep[RGS;][]{herd01}. Each RGS contains 9 CCDs (though one has
failed in each RGS) along the dispersion direction, and is sensitive
to soft X-rays in the range $\sim$ 6--38\unit{\AA}. The field of view
in the cross-dispersion direction is 5\unit{arcmin}.

\subsection{Data reduction}

The two RGS were operated in the standard Spectroscopy mode. Data were
processed from the ODFs using the SAS meta-task \code{rgsproc}. The
position of the cluster emission peak as measured from the MOS image
was used to specify the source extraction point. The generated RGS1
and RGS2 events files were filtered for periods of high background
with the same approach as used in the construction of the RGS
background files \citep{tamu03}. Indeed, since the background files do
not have the necessary information for further filtering, we have
little choice but to proceed in this manner (though we might prefer to
use a stricter filtering in this case). That is, we form a light-curve
from the (\texttt{FLAG == 8,16}) events at high cross-dispersion
angles (\texttt{$|$XDSP\_CORR$| >$ 1.5e-4}) on CCD9 (that which covers
the highest energy band and is most affected by background), using 100
second time bins. Only periods with count-rates less than 0.15 counts
per second for each RGS were retained, leaving $\sim$ 74\unit{ks} of
time per RGS. Observe that although this is significantly larger than
the remaining EPIC exposure, both EPIC and RGS have (essentially) been
filtered to the same level as the appropriate ESA-supplied background
templates (soft protons channelled through the mirrors are responsible
for most of the \xmm{} background flaring, and relatively few are
scattered by the gratings into the RGS detectors).

Since the field of view in the cross-dispersion direction is only
5\unit{arcmin} wide, it is essentially filled by cluster emission from
Abell 2597. We therefore make use of the standard background template
files provided by \citet{tamu03}. We note that these files were made
at an operating temperature of -80\degrees C, but that in late 2002
both RGS were cooled to -110\degrees C.

\begin{figure}
  \centering
  \includegraphics[angle=0,width=0.9\columnwidth]{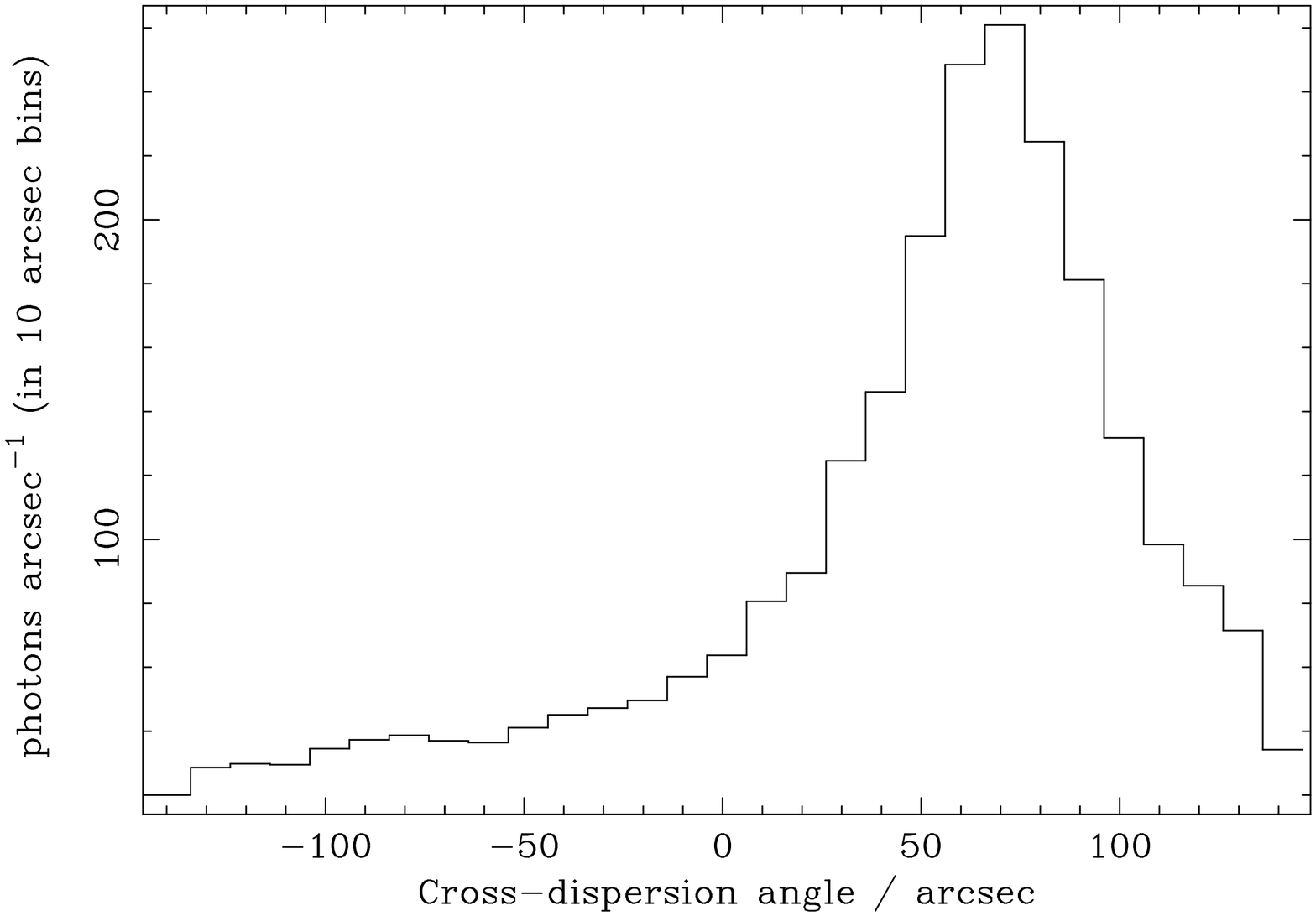}
  \caption{%
    Profile of RGS1 counts in 10\unit{arcsec} bins in the
    cross-dispersion direction, for the events lying in the first
    order PI-$\beta$ (dispersion angle) selection region.
    }%
  \label{fig:rgs1_sb_profile}
\end{figure}

In Fig.~\ref{fig:rgs1_sb_profile} we show the `one-dimensional surface
brightness' profile obtained from RGS1 by summing counts in
10\unit{arcsec} bins in the cross-dispersion direction. The centre of
Abell 2597 is unfortunately offset from the mid-point of the CCD array
by about 1\unit{arcmin}. There are insufficient counts to form
profiles in narrow energy bands, in order to examine the possible
differences between the profiles of different spectral lines, e.g.\
due to resonant scattering \citep{sake02}.

RGS spectra are extracted by making selections both in the spatial
(dispersion/cross-dispersion) and energy (dispersion/PI) planes, by
making use of the intrinsic energy discrimination of the detector
CCDs. Due to the extended nature of the source, we have used slightly
broader selection criteria than the defaults. Our spatial selection
took 97 per cent of the PSF in the cross-dispersion direction, which
is equivalent to the central 2\unit{arcmin} or so of the cluster,
i.e.\ the central cooling region. In the energy plane we took 94 per
cent of the pulse-height distribution. We restricted our analysis to
the first order spectra.

Response files were generated using the SAS task \code{rgsrmfgen} with
standard parameters (4000 energy bins between 0.3 and 2.8\unit{keV}).
Spectra were grouped to a minimum of 20 counts per bin. RGS source and
background spectral channels can have different qualities, and since
\xspec{} ignores quality information in the background spectrum, the
source PHA files were modified by adding in any extra bad channels
from the background spectra. We fit the two RGS instruments
simultaneously over the 5--38\unit{\AA} range, thereby compensating
for the missing iron L complex and \ion{O}{8} line absent (due to
failed CCDs) from RGS1 and RGS2 respectively.

In an attempt to deal with the spectral blurring that results from the
extended nature of the source, we have employed the \xspec{}
\rgsxsrc{} model (due to A. Rasmussen). This convolves the spectral
model with an angular structure function computed from an image (we
used the MOS1 image in the RGS energy range of 0.3--2.5\unit{keV}) of
the target region. Correcting for broadening in this way is only an
approximation to the complex instrument response to a spatially
extended input. The \rgsxsrc{} model assumes that the spatial
structure of the source is independent of energy. To test the possible
consequences of this assumption, we examined the effects of using
input images in different energy ranges (e.g.\ below 1\unit{keV}, or
dividing the RGS energy range in two and using either the upper or
lower half). The results quoted in subsequent sections were found to
have very little dependence on the energy range used for the
\rgsxsrc{} image. For example, the changes in fitted fluxes for the
\ion{Fe}{17} lines (see Section~\ref{sec:rgs_line_fits}) when using a
different energy range for the \rgsxsrc{} image are typically of order
a few per cent, in a random direction (i.e.\ not systematically higher
or lower for an image in any given energy range). The changes in line
significances were negligible.

According to the \textit{\xmm{} Users' Handbook}%
\footnote{\url{http://xmm.vilspa.esa.es/external/xmm_user_support/%
    documentation/uhb/}}, the theoretical wavelength resolution of the
RGS spectral order $n$ for an extended source of angular size $\theta$
(\unit{arcmin}) is degraded according to the formula $\Delta \lambda /
\unit{\AA} \approx 0.138 \; \theta / n$. Thus, for a source region
with extent $\sim 2\unit{arcmin}$ we may expect a wavelength
resolution $\approx 0.25\unit{\AA}$. This is borne out by examination
of the model width of an \rgsxsrc-blurred line of zero intrinsic
width.

\subsection{Broad-band spectral fits}\label{sec:rgs_broad-band_fits}

\begin{figure}
  \centering
  \includegraphics[angle=0,width=1.0\columnwidth]{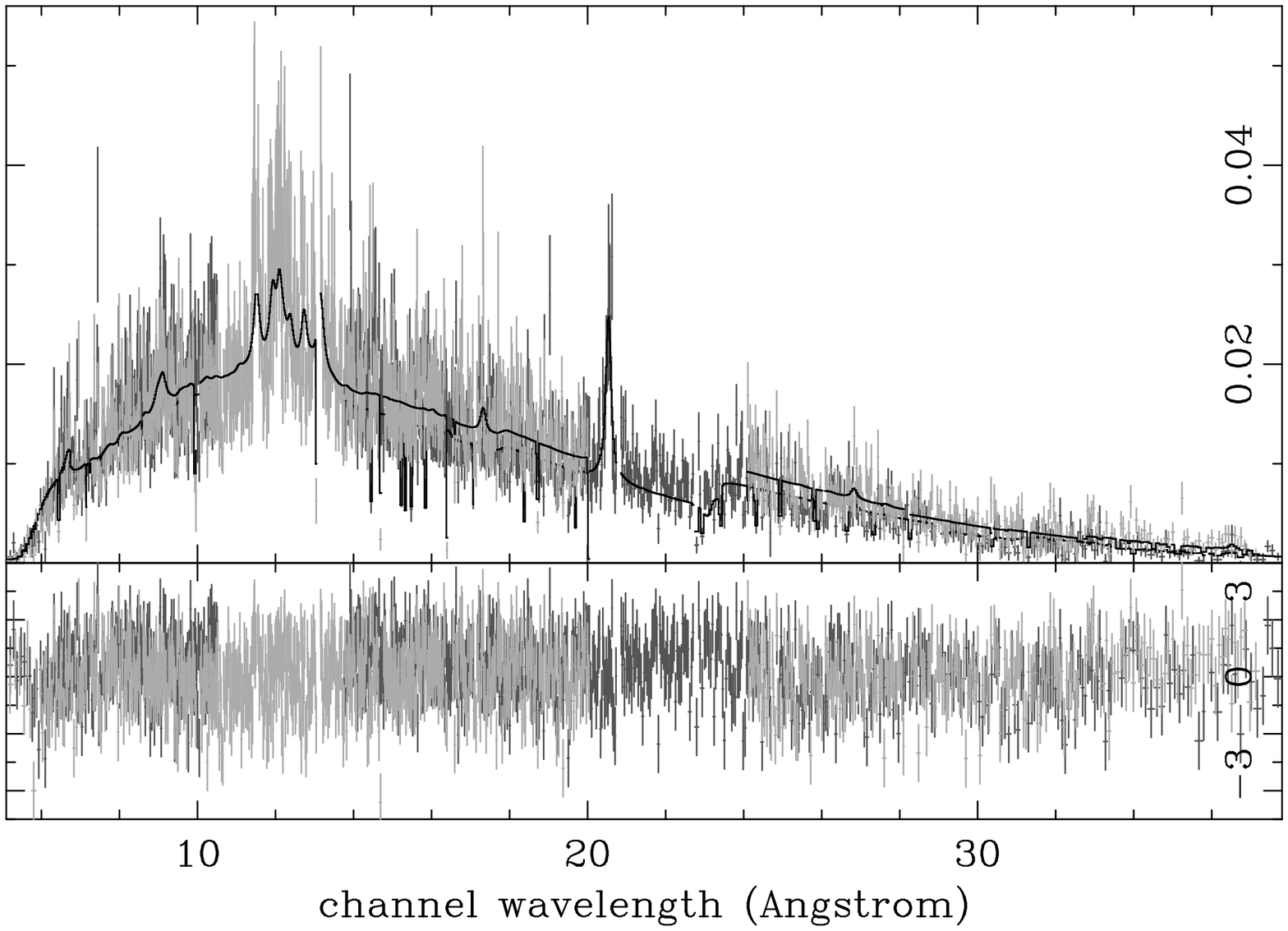}
  \caption{%
    RGS1 (dark) and RGS2 (light) spectra, in units of
    \unit{ct}\unit{\AA^{-1}}\unit{s^{-1}}. Also shown is the
    best-fitting single temperature model, together with its $\chi^{2}$
    residuals (lower panel).}%
  \label{fig:rgs_spec}
\end{figure}

In Fig.~\ref{fig:rgs_spec} we show the best-fitting single temperature
\vmekal{} model and its $\chi^{2}$ residuals as applied to the data.
The abundances of iron and oxygen were allowed to vary independently
(both these elements exhibit strong line features in the RGS energy
range, so we may reasonably expect to obtain reliable constraints on
their individual abundances), whilst those of all the other elements
were tied together. The redshift and absorbing hydrogen column were
also allowed to vary. Good fits (even with a single temperature model)
could not be obtained when the absorption was fixed at Galactic. The
fit parameters are listed in Table~\ref{tab:rgs_fits}. The preferred
absorption is substantially higher than Galactic (and that favoured by
the EPIC instruments), for reasons that are not clear. The excess of
\mysub{N}{H} above Galactic is consistent with the intrinsic
\mysub{N}{H} (Section~\ref{sec:intro}) detected by \citet{odea94}; but
that was only seen against the small central radio source, so this is
probably not significant. Allowing \mysub{N}{H} to vary results in
temperatures and abundances that are consistent with those of the EPIC
fits to the similar spatial region.

\begin{table}
  \centering
  \caption[]{%
    Results of RGS fits to the central $\sim 2$\unit{arcmin} region,
    5--38\unit{\AA}.
  }
  \begin{tabular}{ccccc}\hline
                  & A                            & B                            & C                            & D                            \\\hline
    \mysub{N}{H}  & $6.93_{-0.5}^{+0.2}$         & $6.19_{-0.3}^{+0.3}$         & $6.22_{-0.2}^{+0.2}$         & $6.30_{-0.3}^{+0.2}$         \\
    $kT$          & $2.54_{-0.09}^{+0.16}$       & $3.01_{-0.20}^{+0.13}$       & $3.12_{-0.13}^{+0.06}$       & $3.07_{-0.15}^{+0.22}$       \\
    $kT_{2}$      & -                            & $0.66_{-0.03}^{+0.03}$       & -                            & $0.081_{-0}^{+0.13}$         \\
    $\dot{M}$     & -                            & -                            & $95_{-14}^{+14}$             & $92_{-11}^{+18}$             \\
    \mysub{Z}{O}  & $0.38_{-0.03}^{+0.02}$       & $0.44_{-0.05}^{+0.04}$       & $0.40_{-0.04}^{+0.04}$       & $0.40_{-0.03}^{+0.04}$       \\
    \mysub{Z}{Fe} & $0.22_{-0.02}^{+0.04}$       & $0.35_{-0.05}^{+0.02}$       & $0.34_{-0.03}^{+0.03}$       & $0.33_{-0.02}^{+0.05}$       \\
    \mysub{Z}{x}  & $0.28_{-0.06}^{+0.06}$       & $0.44_{-0.04}^{+0.07}$       & $0.40_{-0.07}^{+0.07}$       & $0.39_{-0.07}^{+0.06}$       \\
    $z$           & $0.0826_{-0.0003}^{+0.0004}$ & $0.0826_{-0.0007}^{+0.0001}$ & $0.0821_{-0.0002}^{+0.0006}$ & $0.0824_{-0.0005}^{+0.0002}$ \\
    $\chi^{2}$    & $1881.2/1670$                & $1852.7/1668$                & $1850.8/1669$                & $1851.5/1668$                \\\hline
  \end{tabular}
  \label{tab:rgs_fits}

  \raggedright

  \mysub{N}{H} is Galactic column in units of
  $10^{20}\unit{cm^{-2}}$; $kT$ temperature in keV; Z metallicity
  relative to solar for O, Fe, and all other elements (x); and $z$
  redshift. Errors are $1\sigma$. All models are \rgsxsrc{} $\times$
  \phabs{} $\times$ (\code{M}), where \code{M} is: A \vmekal; B
  \vmekal{} + \vmekal; C \vmekal{} + \vmcflow{} (with \vmcflow:
  \mysub{T}{max} $\equiv$ \vmekal:$T$, \vmcflow:\mysub{T}{min} $\equiv$
  0.081); D as C but with \mysub{T}{min} free. Where two phases were
  used, the metallicities were tied.
\end{table}

\subsubsection{Redshift issues}\label{sec:rgs_redshift}

Two similar, but distinct, redshifts are in common use for Abell~2597.
\citet{kowa83}, as used in the compilation of \cite{stru99}, measured
the optical redshifts of three (non-cD) cluster galaxies as $z =
0.0874, 0.0832, 0.0851$, hence $\bar{z} = 0.0852 \pm 0.002$ (relative
to Local Group standard of rest). Converting this result to
heliocentric using the NASA/IPAC Extragalactic Database (NED) velocity
converter\footnote{\url{http://nedwww.ipac.caltech.edu/forms/%
    vel_correction.html}} subtracts $\sim 140\unit{km}\unit{s^{-1}}$,
giving $\mysub{z}{helio} = 0.0847$. In contrast, \citet{noon81}, using
\citet{schm65}, has $z = 0.0821$ (heliocentric), based on the central
radio galaxy. \citet{owen95} find a heliocentric $z = 0.0822 \pm
0.0002$, using optical emission lines of the central radio galaxy PKS
2322-12. The best-fit redshift found by \citet{voit97} for the optical
emission line nebula in Abell 2597 is $z = 0.0821 \pm 0.0002$. In
light of all this, we choose to adopt $z = 0.0822$ for the
heliocentric redshift of the cluster.

The best-fitting RGS broad-band redshift is consistent with the
adopted optical redshift. The most prominent single line in the
spectrum is that of \ion{O}{8} \specseries{K}{\alpha} at around
20.5\unit{\AA}. The rest energy of this feature is 653.6\unit{eV}.
Fitting the spectrum in the 19-22\unit{\AA} range with an
\rgsxsrc-blurred power-law plus a (zero intrinsic width) Gaussian, we
find the best-fit value for the energy of this line to be $E =
603.8_{-0.4}^{+0.2}$\unit{eV}, implying a redshift of $z =
0.0825_{-0.0004}^{+0.0007}$, consistent with the broad-band RGS fit,
and the optical value.

We have also fit just the iron L complex, by restricting attention to
the 9.2--17\unit{\AA} range (chosen to exclude the Mg line around
9\unit{\AA}, and to include as much line-free continuum around the Fe
L complex as possible). The plasma properties are not well constrained
in such a narrow wavelength range, with the exception of redshift,
which is our sole interest here. According, we froze \mysub{N}{H} at
the broad-band value, and fit the data with a single \xmekal{} model.
The best-fitting redshift for the Fe L complex was $z =
0.0831_{-0.0016}^{+0.0002}$. As per \citet{stil02}, the best-fit
\ion{O}{8} redshift is formally less than that of the Fe L complex.
Within the 1$\sigma$ limits, however, the redshifts obtained from the
various RGS fits (broad-band, Fe L complex, \ion{O}{8}) are
consistent, and agree with our adopted optical redshift. The
hypothesis of separate velocity components in the core \citep{stil02}
could therefore be supported by these data, but does not seem to be
required.

\subsubsection{Additional fit components}

\begin{figure}
  \centering
  \includegraphics[angle=0,width=0.9\columnwidth]{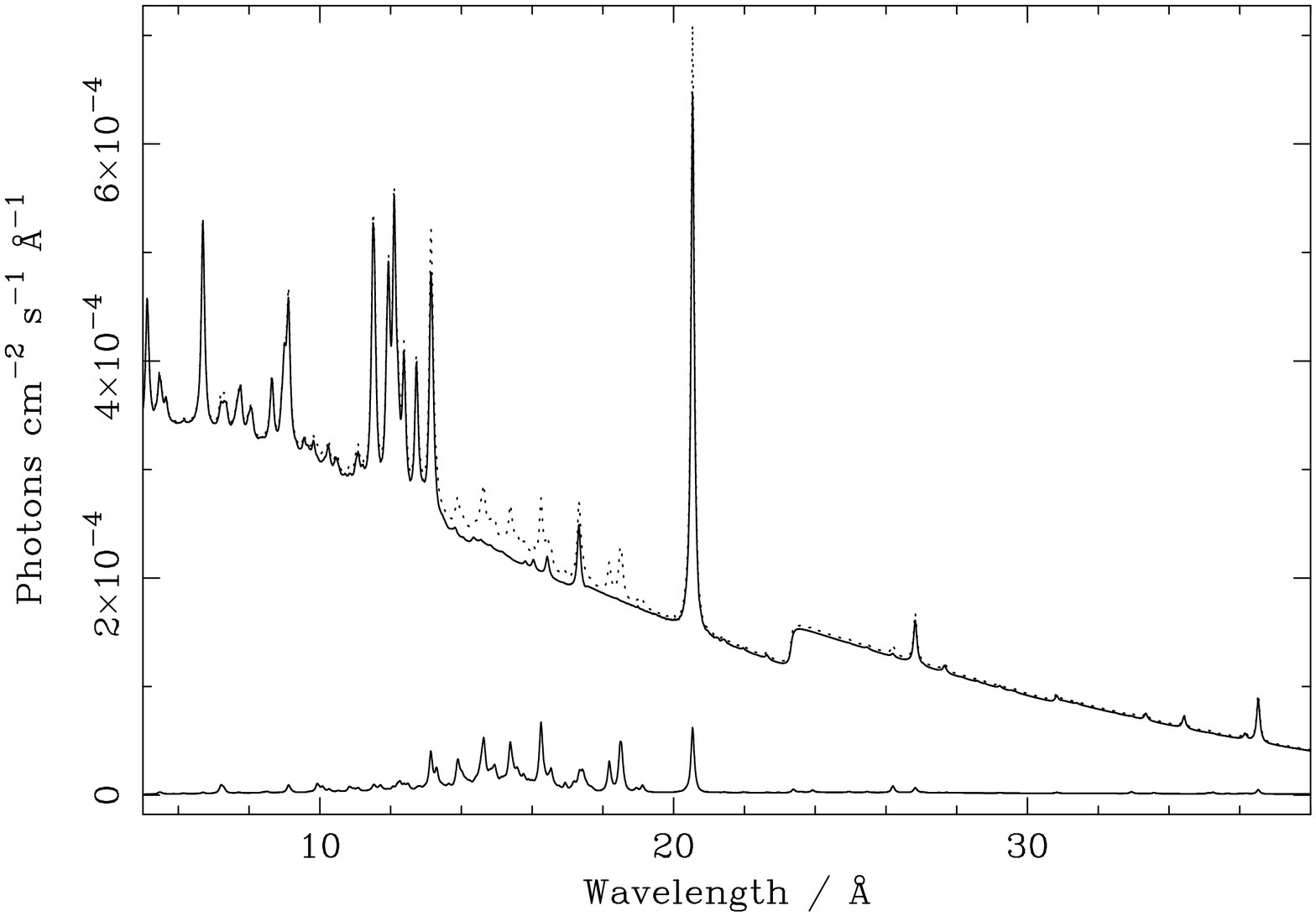}
  \caption{%
    Contributions of the two temperature components (solid lines) of
    model B in Table~\ref{tab:rgs_fits} to the total (dotted line)
    model. The minority component is the low-temperature one.
    }%
  \label{fig:rgs_2mekal_model}
\end{figure}

Also shown in Table~\ref{tab:rgs_fits} are the results of adding a
second temperature component to the fit, with an independent
temperature and normalization. The removal of two degrees of freedom
results in a change in fit of $\Delta \chi^{2} = -28.5$. According to
an F-test, this improvement in fit is significant at the (1 -
\sform{3.1e-6}) level. In Fig.~\ref{fig:rgs_2mekal_model} we show the
contributions made by each of the two temperature components to the
total model. The bulk of the spectral fit is obviously dominated by
the high-temperature component. The low-temperature component
contributes several emission lines at wavelengths around 15\unit{\AA}
-- these are examined in more detail below.

In Table~\ref{tab:rgs_fits} we also show the results of fits using a
cooling flow (\vmcflow{} model) rather than a single temperature for
the second component. These also provide significant improvements to
the single-temperature fit, although the data are not of sufficient
quality to unambiguously state that a cooling-flow fit is preferred.
The most interesting aspect of the fit is the obtained mass deposition
rate: $\dot{M} \sim 100$\unit{\Msun}\unit{yr^{-1}}. This is consistent
with the EPIC fits to the central regions,
Table~\ref{tab:spec_0-4arcmin}.

\begin{figure}
  \centering
  \includegraphics[angle=0,width=0.9\columnwidth]{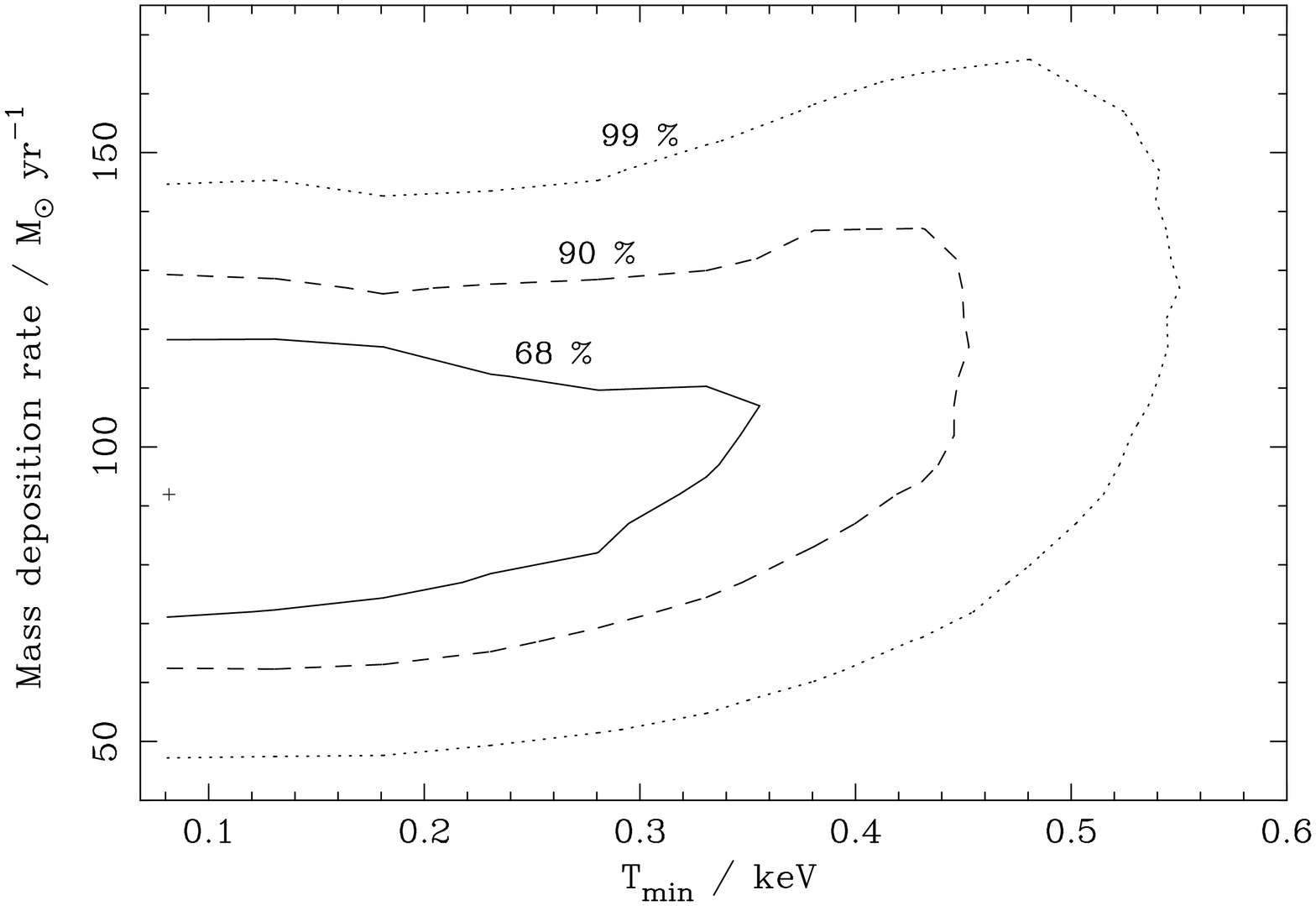}
  \caption{%
    Confidence contours for the RGS cooling flow fit (model D in
    Table~\ref{tab:rgs_fits}) in the $\dot{M}$--\mysub{T}{min} plane.
    Plot details are as per Fig~\ref{fig:mos_contour}. Compare also
    with the \chandra{} results of Fig~\ref{fig:chandra_contour}.
    }%
  \label{fig:rgs_contour}
\end{figure}

Allowing the low-temperature cut-off of the cooling flow component,
\mysub{T}{min}, to be free (rather than fixed at the lowest value
possible in the \xspec{} model, 0.081\unit{keV}), makes no improvement
to the fit quality. Indeed, as we show in Fig.~\ref{fig:rgs_contour},
a very low temperature cut-off for the cooling flow is preferred. The
slight upward curve of the significance contours does indicate that
higher mass deposition rates are permissible (though not preferred)
with higher low-temperature cut-offs.

\subsection{\ion{Fe}{17} line fitting}\label{sec:rgs_line_fits}

We now return to the low-temperature emission lines in the
15\unit{\AA} region of the spectrum previously alluded to. The
\ion{Fe}{17} lines in this region are particularly important because
they are strong indicators of gas at temperatures $\sim$
0.3\unit{keV}, and consequently their presence is an important
prediction of the cooling flow model.

In order to examine these lines in more detail, we fit the
15--20\unit{\AA} region with a blurred power-law and four (zero
intrinsic width) Gaussians. Adopting a fixed redshift of $z = 0.0822$,
the line energies were fixed at the positions appropriate for the
15.02, 16.78, 17.08\unit{\AA} (rest wavelength) \ion{Fe}{17} lines,
and the 16.00\unit{\AA} \ion{O}{8} \specseries{K}{\beta} line. We
found the best-fit and $1\sigma$ limits on the normalizations of each
line. We obtained positive (but weak) evidence for all three important
\ion{Fe}{17} lines, at the 1--2$\sigma$ level (too low to qualify for
a formal detection).

The normalizations were converted to line fluxes (using a luminosity
distance of 503\unit{Mpc}) by subtracting off the continuum component
in the absence of a line. Results are shown in
Table~\ref{tab:rgs_line_fits}.

We have also converted these fluxes to equivalent cooling flow mass
deposition rates. This was achieved by using the \mkcflow{} model in
\xspec{} to make a fake spectrum for a fiducial
1000\unit{\Msun}\unit{yr^{-1}} cooling flow, at a redshift etc.\
appropriate to Abell~2597. We fit the 15--20\unit{\AA} range of this
model spectrum in the same way as applied to the actual data, in order
to obtain the flux expected in each of the \ion{Fe}{17} lines. The
ratios of the Abell~2597 line fluxes to those from the model were used
to estimate the associated mass-flow rates. The results, as shown in
Table~\ref{tab:rgs_line_fits}, are poorly constrained, but consistent
with both the broad-band RGS fits, and the EPIC fits to the central
region.

Table~\ref{tab:rgs_line_fits} also gives the F-test significance (the
comments of Section~\ref{sec:global_props} on the F-test also apply
here) for each of the Fe lines. This is obtained from the improvement
in fit that results when fitting a model with just the \ion{O}{8} and
each Fe line in turn present, compared to the best-fit with just the O
line. The O line has been treated separately due to its different
temperature dependence. The significance of each individual Fe line is
admittedly low, varying from about 1--2$\sigma$. In order to try and
achieve greater significance, we have tied the relative normalizations
of the three \ion{Fe}{17} lines (all with similar temperature
dependences) to that present in the fiducial isobaric cooling flow
model, and just allowed the overall normalization to vary. The results
are given in the `All' row of Table~\ref{tab:rgs_line_fits}. When
combining the lines in this way, the significance is still low, but
exceeds $2\sigma$.

\begin{table}
  \centering
  \caption[]{%
    RGS \ion{Fe}{17} line fits.
  }
  \begin{tabular}{cccc}\hline
    $\lambda / \unit{\AA}$ &
    Power / $10^{40}$\unit{erg}\unit{s^{-1}} &
    $\dot{M}$ / \unit{\Msun}\unit{yr^{-1}} & \% \\\hline
    15.02 & $14_{-14}^{+15}$ & $34_{-34}^{+37}$  & 54 \\
    16.78 & $19_{-14}^{+12}$ & $100_{-71}^{+63}$ & 86 \\
    17.08 & $23_{-14}^{+12}$ & $57_{-34}^{+31}$  & 91 \\
    All   & $52_{-23}^{+24}$ & $52_{-23}^{+24}$  & 97 \\\hline
  \end{tabular}
  \label{tab:rgs_line_fits}

  \raggedright
  Column~1: rest wavelength; column~2: power in line ($\mysub{D}{L} =
  503\unit{Mpc}$);
  column~3: mass-flow rate in line, inferred by comparison with a
  fiducial isobaric cooling flow model (see text for details);
  column~4: F-test significance (see text).
\end{table}

\begin{figure}
  \centering
  \includegraphics[angle=0,width=1.0\columnwidth]{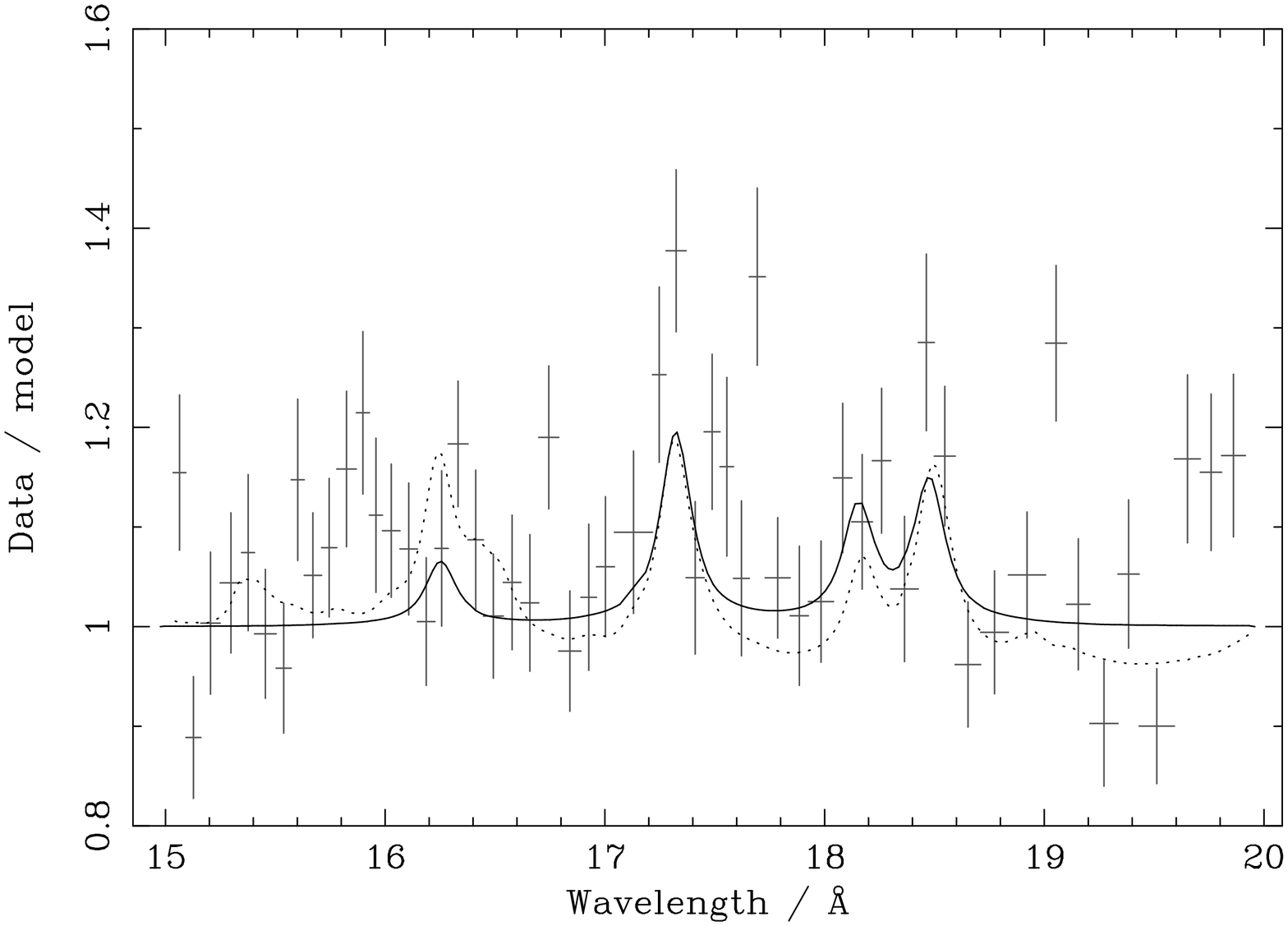}
  \caption{%
    Ratio of data to a simple power-law model (i.e.\ no emission
    lines) in the 15--20\unit{\AA} region. The data have been
    re-binned, and the two RGS instruments have been combined (this
    requires a degree of interpolation since the wavelengths bins in
    the two instruments do not coincide exactly), for display purposes
    only. The solid line shows the ratio appropriate for the Gaussian
    emission lines fitted in Table~\ref{tab:rgs_line_fits}; the dotted
    line that for the cooling flow model C of
    Table~\ref{tab:rgs_fits}. Note that the models were fitted to the
    un-binned data for the two separate instruments simultaneously
    (i.e.\ the models were not fitted to the data as displayed here).
  }%
    \label{fig:rgs_ratio}
\end{figure}

In Fig.~\ref{fig:rgs_ratio} we show the 15--20\unit{\AA} region of
the RGS spectrum, in terms of the ratio of the data to a model with no
emission lines. Also shown are the appropriate curves for the model
fitted for Table~\ref{tab:rgs_line_fits} (power-law plus Gaussians)
and for model C of Table~\ref{tab:rgs_fits} (isobaric cooling flow).
We note that the normalizations adopted by the Gaussians (where the
line strengths may vary independently) are very similar to those of
the cooling flow model. The exception is the 15.02 line, which is the
\ion{Fe}{17} line most susceptible to resonant scattering
\citep[e.g.][see also the APED
database\footnote{\url{http://cxc.harvard.edu/atomdb/}}]{rugg85},
for which a weaker fit is preferred. The quality of the line fits is
clearly poor, but the values obtained for mass deposition rates agree
with those from broad-band fits in both EPIC and the RGS.

\section{Discussion}

First, we must caution that the overall quality of our observation is
low, and that the formal statistical significance of our results is
not large. In particular, the results on the \ion{Fe}{17} lines are
really no more than suggestive. Nevertheless, the data do present a
self-consistent picture between the various detectors, with an
extremely interesting interpretation.

Fitting each of the three EPIC detectors individually for a large
central region covering that radial range where the cluster emission
is high, then the addition of a cooling flow component provides a
highly significant improvement in fit over a single temperature model
(although we cannot statistically distinguish a cooling flow
component, with a range of temperatures, from a single temperature
second component). The best-fitting mass deposition rate is around
90\unit{\Msun}\unit{yr^{-1}}, with a $1\sigma$ error of about
15\unit{\Msun}\unit{yr^{-1}} (see Table~\ref{tab:spec_0-4arcmin} for
details). The cooling time is less than 10\unit{Gyr} within a radius
of 130\unit{kpc} (Fig.~\ref{fig:ne_tcool_S_deproj}).

Very similar results are obtained from the broad-band simultaneous fit
to the two RGS detectors (Table~\ref{tab:rgs_fits}). The main evidence
in the data for low-temperature gas comes from the spectral region
around 15\unit{\AA}, in particular the emission lines of \ion{Fe}{17}
at 15, 17\unit{\AA} rest wavelength. We obtain weak (1--2$\sigma$;
below formal detection levels) evidence for the three most important
lines, and measure the power associated with each line (see
Table~\ref{tab:rgs_line_fits}). By comparing the line fluxes with
those produced from a classical isobaric cooling flow model, we can
estimate the corresponding mass-flow rate for each line. Once again,
we obtain values that are consistent with the
90\unit{\Msun}\unit{yr^{-1}} mark. Fixing the relative strengths of
the three Fe lines to that predicted by the standard isobaric cooling
flow model, and allowing the absolute normalization to vary, we obtain
a fit suggesting the presence of \ion{Fe}{17} at a significance of
just over $2\sigma$.

We therefore establish, using three distinct methods, evidence for a
cooling flow at a level $\sim 90\unit{\Msun}\unit{yr^{-1}}$. The
preferred low-temperature cut-off for the flow, established from the
RGS data, is essentially zero (0.081\unit{keV}, the minimum
temperature available in the \xspec{} \xmekal{} model). A plot of the
$\dot{M}$--\mysub{T}{min} confidence contours is shown in
Fig.~\ref{fig:rgs_contour}.

A re-analysis of the \chandra{} data for Abell~2597 \citep{mcna01}, as
described in Section~\ref{sec:chandra_global}, also supports the
existence of a cooling flow at these levels, with a very small
low-temperature cut-off -- see Fig.~\ref{fig:chandra_contour}.

This mass deposition rate is consistent with the results of
\citet{oege01}, using \fuse{} UV observations. These authors detected
the \ion{O}{6} 1032\unit{\AA} resonance line (characteristic of gas at
temperatures $\sim$ \sform{3e5}\unit{K}) in Abell~2597. Converting the
detected flux to an equivalent mass deposition rate gives a UV
mass-deposition rate $\sim 40\unit{\Msun}\unit{yr^{-1}}$ within the
\fuse{} effective radius of 40\unit{kpc}. This is consistent with the
X-ray results (Fig.~\ref{fig:mdot_cumu}).

In contrast, \ion{O}{6} was not detected in Abell~1795 \citep{oege01},
nor in Abell~2029 or Abell~3112 \citep{leca04}. Both from an X-ray and
a UV standpoint, therefore, Abell~2597 appears to be an atypical
object. The combination of the \xmm{} X-ray and \fuse{} UV results for
Abell~2597 suggests that it may harbour a classical cooling flow in
which gas cools from $\sim$ 4\unit{keV} by at least two orders of
magnitude in temperature. Interestingly, a recent \chandra{} analysis
of Abell~2029 \citep{clar04} shows that the X-ray data in this object
are also consistent with a modest (by traditional standards) cooling
flow extending down to very low temperatures, despite the lack of a UV
detection.

Among the many \citep[e.g.][]{fabi01a,pete01} explanations suggested
for the `cooling-flow problem', heating from a central AGN, mediated
by the rise of buoyant plasma bubbles through the ICM, is a popular
candidate \citep[e.g.][]{vecc04,reyn05,rusz04}. The \chandra{}
observation of Abell~2597 \citep{mcna01} showed it to contain X-ray
surface brightness depressions, interpreted (as a result of the
extension of spurs of old, low-frequency radio emission) as `ghost
cavities', associated with a radio outburst $\sim 10^{8}$\unit{yr}
ago.

Abell~2597 therefore appears to harbour a central AGN that generates
buoyant bubbles, and yet it also appears to contain gas cooling to
very low temperatures. The process of bubble generation is necessarily
an episodic one, and it therefore perhaps unreasonable to assume that
a steady, time-independent state will be maintained. Clusters may go
through cycles of behaviour, in which cooling builds up to some level,
then an AGN outburst initiates a heating phase in which cooling is
suppressed (else massive cooling flows would be common phenomena).
Abell~2597 would then be an object near the `cooling catastrophe'
point of the cycle. The relative frequency of such objects would
depend on the relative timescales of the cooling and heating phases.
The fact that such objects seem to be quite rare indicates that the
phase of strong cooling is short-lived.

Such a mechanism would require coupling between the state of the ICM
and the activity of the central AGN, in order that a feedback loop
could be maintained.

\section*{Acknowledgments}

This work is based on observations obtained with \xmm, an ESA science
mission with instruments and contributions directly funded by ESA
Member States and NASA. We thank PPARC (RGM) and the Royal Society
(ACF) for support. RGM is grateful to Steve Allen for helpful
discussions; to Jeremy Sanders for some useful \xspec{} scripts; and
to Keith Arnaud for tireless assistance with the \xmmpsf{} model.

\bibliographystyle{mnras}

\end{document}